\documentclass[twocolumn]{aastex61}

\newcommand{\hi}{\ion{H}{1}}
\received{Feb 22, 2018}
\submitjournal{ApJS}

\begin{document}

\title{An H$\alpha$ Imaging Survey of the Low$-$surface$-$brightness Galaxies Selected from the Fall Sky Region of the 40$\%$ ALFALFA \ion{H}{1} Survey}

\correspondingauthor{ Hong Wu}
\email{hwu@bao.ac.cn}

\author{Feng$-$Jie Lei}
\affiliation{Nationl Astronomical Observatories, Chinese Academy of Sciences, 20A Datun Road, Chaoyang District, Beijing, 100012, China}
\affiliation{Key Laboratory of Optical Astronomy, National Astronomical Observatories, Chinese Academy of Sciences, Beijing 100012, P.R. China}
\affiliation{School of Astronomy and Space Science University of Chinese Academy of Sciences, Beijing 100049, China; fjlei@nao.cas.cn}

\author{Hong Wu}
\affiliation{Nationl Astronomical Observatories, Chinese Academy of Sciences, 20A Datun Road, Chaoyang District, Beijing, 100012, China}
\affiliation{Key Laboratory of Optical Astronomy, National Astronomical Observatories, Chinese Academy of Sciences, Beijing 100012, P.R. China}
\affiliation{School of Astronomy and Space Science University of Chinese Academy of Sciences, Beijing 100049, China; fjlei@nao.cas.cn}

\author{Wei Du}
\affiliation{Nationl Astronomical Observatories, Chinese Academy of Sciences, 20A Datun Road, Chaoyang District, Beijing, 100012, China}
\affiliation{Key Laboratory of Optical Astronomy, National Astronomical Observatories, Chinese Academy of Sciences, Beijing 100012, P.R. China}

\author{Yi$-$Nan Zhu}
\affiliation{Nationl Astronomical Observatories, Chinese Academy of Sciences, 20A Datun Road, Chaoyang District, Beijing, 100012, China}
\affiliation{Key Laboratory of Optical Astronomy, National Astronomical Observatories, Chinese Academy of Sciences, Beijing 100012, P.R. China}

\author{Man$-$I Lam}
\affiliation{Shanghai Astronomical Observatories, Chinese Academy of Sciences, 80 Nandan Road, Xuhui District, Shanghai, 200030, China}

\author{Zhi$-$Min Zhou}
\affiliation{Nationl Astronomical Observatories, Chinese Academy of Sciences, 20A Datun Road, Chaoyang District, Beijing, 100012, China}
\affiliation{Key Laboratory of Optical Astronomy, National Astronomical Observatories, Chinese Academy of Sciences, Beijing 100012, P.R. China}

\author{Min He}
\affiliation{Nationl Astronomical Observatories, Chinese Academy of Sciences, 20A Datun Road, Chaoyang District, Beijing, 100012, China}
\affiliation{Key Laboratory of Optical Astronomy, National Astronomical Observatories, Chinese Academy of Sciences, Beijing 100012, P.R. China}
\affiliation{School of Astronomy and Space Science University of Chinese Academy of Sciences, Beijing 100049, China; fjlei@nao.cas.cn}

\author{Jun$-$Jie Jin}
\affiliation{Nationl Astronomical Observatories, Chinese Academy of Sciences, 20A Datun Road, Chaoyang District, Beijing, 100012, China}
\affiliation{Key Laboratory of Optical Astronomy, National Astronomical Observatories, Chinese Academy of Sciences, Beijing 100012, P.R. China}
\affiliation{School of Astronomy and Space Science University of Chinese Academy of Sciences, Beijing 100049, China; fjlei@nao.cas.cn}

\author{Tian$-$Wen Cao}
\affiliation{Nationl Astronomical Observatories, Chinese Academy of Sciences, 20A Datun Road, Chaoyang District, Beijing, 100012, China}
\affiliation{Key Laboratory of Optical Astronomy, National Astronomical Observatories, Chinese Academy of Sciences, Beijing 100012, P.R. China}
\affiliation{School of Astronomy and Space Science University of Chinese Academy of Sciences, Beijing 100049, China; fjlei@nao.cas.cn}

\author{Pin$-$Song Zhao}
\affiliation{Nationl Astronomical Observatories, Chinese Academy of Sciences, 20A Datun Road, Chaoyang District, Beijing, 100012, China}
\affiliation{Key Laboratory of Optical Astronomy, National Astronomical Observatories, Chinese Academy of Sciences, Beijing 100012, P.R. China}
\affiliation{School of Astronomy and Space Science University of Chinese Academy of Sciences, Beijing 100049, China; fjlei@nao.cas.cn}

\author{Fan Yang}
\affiliation{Nationl Astronomical Observatories, Chinese Academy of Sciences, 20A Datun Road, Chaoyang District, Beijing, 100012, China}
\affiliation{Key Laboratory of Optical Astronomy, National Astronomical Observatories, Chinese Academy of Sciences, Beijing 100012, P.R. China}

\author{Chao$-$Jian Wu}
\affiliation{Nationl Astronomical Observatories, Chinese Academy of Sciences, 20A Datun Road, Chaoyang District, Beijing, 100012, China}
\affiliation{Key Laboratory of Optical Astronomy, National Astronomical Observatories, Chinese Academy of Sciences, Beijing 100012, P.R. China}

\author{Hong$-$Bin Li}
\affiliation{Nationl Astronomical Observatories, Chinese Academy of Sciences, 20A Datun Road, Chaoyang District, Beijing, 100012, China}
\affiliation{Key Laboratory of Optical Astronomy, National Astronomical Observatories, Chinese Academy of Sciences, Beijing 100012, P.R. China}

\author{Juan$-$Juan Ren}
\affiliation{Nationl Astronomical Observatories, Chinese Academy of Sciences, 20A Datun Road, Chaoyang District, Beijing, 100012, China}
\affiliation{Key Laboratory of Optical Astronomy, National Astronomical Observatories, Chinese Academy of Sciences, Beijing 100012, P.R. China}

\begin{abstract}
We present the observed H$\alpha$ flux and derived star formation rates (SFRs) for a fall sample of low$-$surface$-$brightness galaxies (LSBGs).
The sample is selected from the fall sky region of the 40$\%$ ALFALFA {\ion{H}{1}} survey $-$ SDSS DR7 photometric data, and all the $H\alpha$ images were obtained using 
the 2.16 m telescope, operated by the National Astronomy Observatories, Chinese Academy of Sciences. 
A total of 111 LSBGs were observed and $H\alpha$ flux was measured in 92 of them.
Though almost all the LSBGs in our sample are {\ion{H}{1}}$-$rich, 
their SFRs derived from the extinction and filter$-$transmission$-$corrected $H\alpha$ flux, are less than 1$M_{\sun}$$yr^{-1}$.
 LSBGs and star forming galaxies have similar {\ion{H}{1}} surface densities, but LSBGs have much lower SFRs and SFR surface densities than star$-$forming galaxies. 
Our results show that LSBGs deviate from the Kennicutt-Schmidt law significantly, which indicate that they have low star formation efficiency.

The SFRs of LSBGs are close to average SFRs in Hubble time and support the previous arguments that most of the LSBGs are stable systems and they tend to seldom contain strong interactions or major mergers during their star formation histories.

\end{abstract}

\section{Introduction} \label{sec:intro}
Low$-$surface$-$brightness galaxies (LSBGs) are galaxies whose central surface brightness is at least one magnitude fainter than the level of sky background in the dark night\citep{1970ApJ...160..811F,1997ARA&A..35..267I}. 
Generally, they are defined as central surface brightness in the B$-$band $\mu_{0}(B)$ $>$  22.0-23.0 $mag\ arcsec^{-2}$\citep{2001AJ....122.2341I,2012MNRAS.426L...6C}. 
LSBGs account for the bulk of the number of local galaxies, making them an important contributor to the baryon and dark matter mass budget in the local universe \citep{2000ApJ...529..811O,2005ApJ...631..208B,2016A&A...593A.126B}.
Their morphologies and stellar populations distribute widely, ranging from old, high-metallicity early types to young, low-metallicity late-type galaxies \citep{2000MNRAS.312..470B}.
Even though the specific procedure of their formation and evolution is still unclear, their lower star formation rate (SFR) is consistent with the hypothesis that they are quiescent galaxies and have different star formation histories from their high surface brightness counterparts \citep{1995AJ....109.2019M,1999A&A...342..655G,2008ApJ...681..244B,2009ApJ...696.1834W,2013AJ....146...41S}.

One of the most important parameters for understanding the evolution of galaxies is SFR.
There are many approaches to deriving the SFR, utilizing the luminosity related to young massive stars, such as H$\alpha$, UV, or IR luminosities, or fitting the observed spectral energy distribution with a model \citep{1998ARA&A..36..189K,1998ApJ...509..103S,2005ApJ...632L..79W,2008MNRAS.388.1595D,2008ApJ...686..155Z,2009A&A...507.1793N,2009ApJ...706.1527B,2014MNRAS.438...97W,2016ApJ...825...34J}.
Among those SFR tracers, H$\alpha$ emission is connected with the photons whose wavelengths are shorter than the 912{$\rm \AA$}.  
These ionized photons are produced by young stars with ages of less than $\sim$10 $Myr$ and masses higher than 17 {$\rm M_{\sun}$} \citep{2016MNRAS.455.1807W}. 
Therefore, compared to the approaches, the star formation timescale traced by H$\alpha$ emission is shorter.

Recent and ongoing H$\alpha$ image surveys provide a number of resources to study star formation.
The H$\alpha$3 survey is an H$\alpha$ narrow band imaging survey of the Local and Coma Super$-$clusters selected from ALFALFA  \citep{2011AJ....142..170H}, which present the complete recent star formation and {\hi}$-$rich galaxies in the Local Supercluster \citep{2013A&A...553A..89G,2013A&A...553A..91F}.
\citet{2016ApJ...824...25V} finished observations and data reduction for a fall sample of 656 galaxies from the {\hi} Arecibo Legacy Fast ALFA Survey (ALFALFA), the galaxies distances between $\sim$20 and $\sim$100 Mpc, but there was not focus on LSBGs. 
There is an ongoing H$\alpha$ image survey of LSBGs selected from the PSS-II catalog \citep{1992AJ....103.1107S}. 
However, only 59 LSBGs have been included in \citet{2011AdAst2011E..12S}$'s$ sample. 
Consequently, up to now, there are only a few H$\alpha$ surveys of LSBGs, and the total number of LSBGs with available H$\alpha$ photometry is not large enough to derive confirming results.
Therefore, we undertake an H$\alpha$ survey to follow up {\hi}$-$selected LSBGs Galaxies from the 40\% ALFALFA {\hi} survey \citep{2015AJ....149..199D}, 
and we aim to study the SFR and star formation efficiency(SFE) of the {\hi}$-$selected LSBGs.

There is an empirical relation between the gas surface density($\Sigma_{gas}=\Sigma_{HI+H_{2}}$) and SFR surface density ($\Sigma_{SFR}$), ($\Sigma_{SFR}\varpropto\Sigma_{gas}^{1.4}$). Known as the Kennicutt-Schmidt Law, it reflects the relation between the large$-$scale SFR and the physical conditions in the interstellar medium \citep{1959ApJ...129..243S,1998ApJ...498..541K,2008AJ....136.2846B,2008AJ....136.2782L,2008ApJ...681..244B,2010AJ....140.1194B,2009ApJ...696.1834W,2013ApJ...769L..12L,2016A&A...593A.126B}.
However, such an empirical relation, generally derived on the basis of the samples of normal galaxies, might not be suitable for dwarf galaxies or LSBGs \citep{2012AJ....143..133H}.
 \citet{2011ApJ...733...87S} proposed an "extended Schmidt Law," which can be suitable for LSBGs.

In this paper, we present an H$\alpha$ survey for a sample of 111 LSBGs in the fall season in order to explore their SFRs and SFEs.
This paper is orgnized as follows. in Section 2, we introduce our sample together with a description of the observations and data reduction.
In section 3, we present the catalog of H$\alpha$ flux and some derived parameters.
Results and an analysis are given in section 4, and a summary is provided in section 5.
Throughout the paper we adopt a flat $\Lambda$CDM cosmology, with $H_{0}$ = 70 km\ $s^{-1}Mpc^{-1}$ and $\Omega_{\Lambda}$ = 0.7.

\section{SAMPLE OBSERVATIONS AND DATA REDUCTION}
\subsection{Sample}
The ALFALFA Survey is a second-generation blind extragalactic {\hi} survey and provides the first full census of {\hi}-bearing objects over a cosmologically significant volume in the local Universe.
This extragalactic {\hi} survey is especially useful for studying low-mass, gas-rich objects in the local universe \citep{2005AJ....130.2598G,2011AJ....142..170H,2014ApJ...793...40H}.
 This survey covers 7000 {$\rm deg^2$} and intends to detect more than 30,000 extragalactic {\hi} sources.
 The first release covers 40\% of the ALFALFA survey area and is called $\alpha$.40 \citep{2011AJ....142..170H}.

\citet{2015AJ....149..199D} constructed an LSBGs sample with {$\rm \mu_0(B)>22.5$ $mag\ arcsec^{-2}$} from ALFALFA $\alpha.40$ in conjunction with SDSS DR7 photometry data \citep{2009ApJS..182..543A} with an additional constraint on the axis ratio($b/a>0.3$) to prevent the contamination from the edge-on galaxies.
Because the SDSS pipeline overestimates the level of sky background and underestimates the total magnitude of galaxies by about 0.2 mag, this value can reach 0.5 mag for LSBGs \citep{2007ApJ...660.1186L,2013ApJ...773...37H}.
\citet{2015AJ....149..199D} reconstructed the sky background with a better method \citep{1999AJ....117.2757Z,2002AJ....123.1364W,2013ApJ...773...37H} to get more accurate surface brightness. The galaxy geometric parameters (e.g., disk scale length in pixels, axis ratio) are fitted and obtained by software GALFIT \citep{2002AJ....124..266P} and central surface brightness in g-bnd and r-band are calculated by auto-magnitudes from the software Sextractor \citep{1996A&AS..117..393B}.
The central surface brightnesses in B-band are transformed from SDSS g- and r-band magnitudes. The final sample includes 1129 {\hi}$-$rich LSBGs, which are defined as the main LSBG sample; hereafter they are referred to as Du2015.

Our sample contains fall objects (111) from Du2015 (1129) 
and is located within the region of  { $\rm 21^{h} < R.A. < 2^{h} ; $ $\rm 13^{\circ} < Dec. < 16^{\circ}$ and $\rm 23^{\circ} < Dec. < 33^{\circ}$}.
 To obtain more accurate SFRs of LSBGs, an H$\alpha$ imaging survey is needed.
We observed the H$\alpha$ images of a sample of 111 LSBGs located in the fall sky. 
 All members of our LSBGs sample are belong to a blue cloud and are in a star formation sequence.

\begin{figure*}
\epsscale{0.7}
\plotone{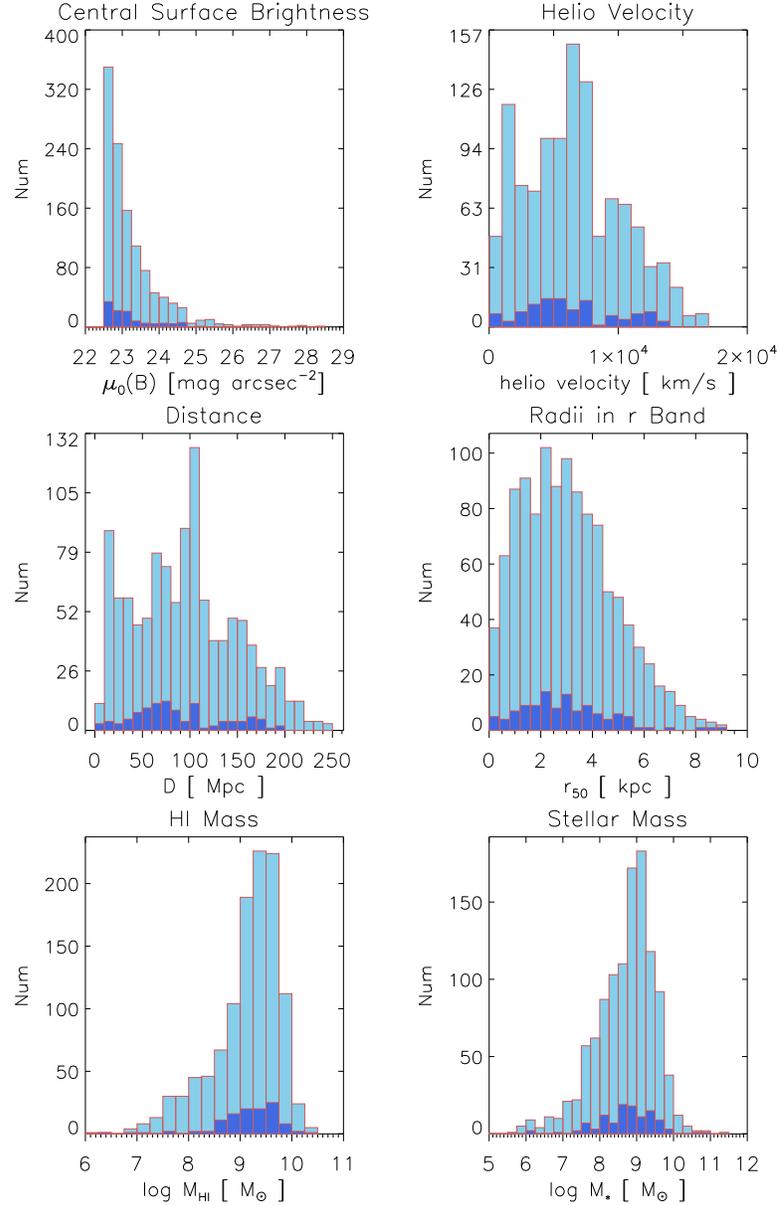}
\caption{ Photometric and {\hi} parameters of our sample(royal blue) and the 
entire LSBG sample(sky blue) \citep{2015AJ....149..199D} . 
The parameters are $B$-band central surface brightness with a bin size of 0.25 mag (top left), 
heliocentric velocity with a bin size of 1000 km$\thinspace s^{-1}$ (top right),
distance with a bin size of 10 Mpc (middle left),
 radius containing 50$\%$ of flux($r_{50}$) in the SDSS $r$-band derived from \citet{2015AJ....149..199D} (middle right),
 {\hi} mass with a bin size of 0.25 (bottom left),
and stellar mass derived from $g$- and $r$-band magnitude with bin sizes of 0.25 (bottom right).
 }
\label{fig1}
\end{figure*}

We show the distributions of some photometric and {\hi} parameters, including central surface brightness, heliocentric velocity, distance, radius containing 50$\%$ of Petrosian flux($r_{50}$) in the SDSS r-band, {\hi} mass, and stellar mass, of the LSBGs in our fall sample (royal blue) and Du2015(sky blue) in Figure \ref{fig1}.
All the {\hi} parameters (heliovelocity, distance, {\hi} mass) are derived from the $\alpha$.40 catalog, and the heliocentric velocity of the {\hi} source $cz_{\sun}$ is in units of km$\thinspace$$s^{-1}$ \citep{2011AJ....142..170H}.

Central surface brightness and $r_{50}$ and g,r magnitudes are from Du2015.
The stellar mass is derived from the $r$-band magnitude and the $g-r$ color using the formula from \citet{2003ApJS..149..289B}.

 The distances used in this paper are estimated from two different approaches\citep{2011AJ....142..170H}.
when the recession velocity (c$z_{\odot}$) of a galaxy is larger than 6000 km$\thinspace$$s^{-1}$, the distance is estimated from c$z_{cmb}$$/$$H_{0}$;  for those whose c$z_{\odot}$ $<$ 6000 km$\thinspace$$s^{-1}$, a velocity model is used \citep{2011AJ....142..170H} to derive their distances.
The peak of the {\hi} mass distribution of our sample is  {$ \rm logM_{HI}[M_{\odot}] \thicksim 9.7$}.
According to \citet{2014ApJ...793...40H} classification,
30$\%$ of LSBGs have high {\hi} mass {($\rm log M_{HI}[M_{\odot}] \geqslant 9.5$ )}, 65$\%$ LSBGs have medium {\hi} mass {(7.7 $\rm  \leqslant log M_{HI}[M_{\odot}] \leqslant 9.5$)}, and only 5$\%$ of LSBGs have low {\hi} mass {(\rm $log M_{HI}[M_{\odot}] \leqslant 7.7$)}.
The peak of the stellar mass is around $10^{8.5}-10^{9}$ $[M_{\odot}]$ .

\subsection{Observation}
The observation for this LSBG sample ranged from 2014 to 2016, and the galaxies in our sample were taken in dark night.
Both broad $R$-band and H$\alpha$ narrow band images were obtained with the BAO Faint Object Spectrograph and Camera (BFOSC) attached to the 2.16 m telescope at Xinglong observatory of the National Astronomical Observatories, Chinese Academy of Sciences (NAOC).
The CCD frame of BFOSC is 1152$\times$1274 $pixel^2$ with the pixel scale of 0.45 $arcsec$ and has a field of view (FOV) of 8.5$\times$9.5 $arcmin^2$. 
The observation was made with a gain mode of 1.08 $e^-$ $\mathrm{ADU^{-1}}$ with a readout noise of 3 $e^-$ $pixel^{-1}$ .
The FOV is suitable for acquiring the images of galaxies with sizes of less than 3-4 arcmin, owing to the fact that the accurate estimation the of sky background is essential for LSBGs.

\begin{figure*}
\epsscale{0.8}
\plotone{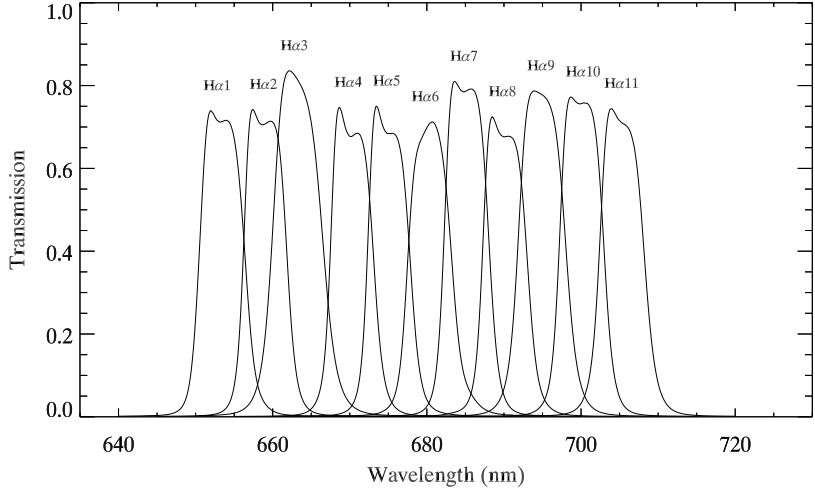}
\caption{ Transmission curve of $\rm H\alpha$ filters from $\rm H\alpha$1 to $\rm H\alpha$11.  
 }
\label{fig2}
\end{figure*}

Each observation adopts the same $R$-band filter and a suitable $H\alpha$ filter.
The effective wavelength $\lambda_{\mathrm{eff}}$ of the broad $R$-band filter is 6407$\rm \AA$ .
There is a series of narrow band H$\alpha$ filters whose center wavelengths range from 6533 to 7052 $\rm \AA$ (6533, 6589, 6631, 6701, 6749, 6804, 6851, 6900$\rm \AA$ and 6948, 7000, and 7052$\rm \AA$) with an FWHM of about 55 $\rm \AA$.
All the central wavelengths and FWHMs of the $H\alpha$ filters are shown in Table \ref{tab:table1}.
The transmission curves of narrow $H\alpha$ filters are shown in Figure \ref{fig2}.

\begin{deluxetable}{ccc}
\tablecolumns{3}
\tablewidth{-20pt}
\tablecaption{The Properties of $\rm H\alpha$ narrow band filters\label{tab:table1}}
\tablehead { \colhead{Filter}&  \colhead{$\lambda_{c}$} & \colhead{FWHM} \\
\colhead{} &  \colhead{$\rm \AA$} & \colhead{$\rm \AA$} \\
\colhead{(1)} &\colhead{(2)} &\colhead{(3)} \\ }
\startdata
  $\rm H\alpha1$   &   6533  &   55  \\
  $\rm H\alpha2$   &   6589  &   53  \\
  $\rm H\alpha3$   &   6631  &   62  \\
  $\rm H\alpha4$   &   6701  &   53  \\
  $\rm H\alpha5$   &   6749  &   52  \\
  $\rm H\alpha6$   &   6804  &   54  \\
  $\rm H\alpha7$   &   6851  &   54  \\
  $\rm H\alpha8$   &   6900  &   55  \\
  $\rm H\alpha9$   &   6948  &   58  \\
  $\rm H\alpha10$  &   7000  &   54  \\
  $\rm H\alpha11$  &   7052  &   56  \\
\enddata
\end{deluxetable}

For each source, the $R$ and H$\rm \alpha$ images were taken with exposures of 300s ($R$) and 1800s ($H\alpha$ narrow band), respectively.  
The $R$-band integration time is deep enough to provid continuum subtraction for the narrow band image.
The observation information is listed in Table \ref{table2}.

\subsection{Image Reduction}

Firstly, we check the quality of the images with the naked eye. After that, we reduce the CCD frames, including overscan subtraction, bias subtraction, flat-field correction, and cosmic-ray removal, following the standard image process with IRAF provided 
by NOAO \footnote{IRAF is the Image Analysis and Reduction Facility made available to the astronomical community by the National Optical Astronomy Observatories, which are operated by AURA, Inc., under contract with the US National Science Foundation. STSDAS is distributed by the Space Telescope Science Institute, which is operated by the Association of Universities for Research in Astronomy (AURA), Inc., under NASA contract NAS 5–26555.}
Then, the celestial coordinates are added to each image using $Astrometry.net$.

\begin{figure*}
\epsscale{0.8}
\plotone{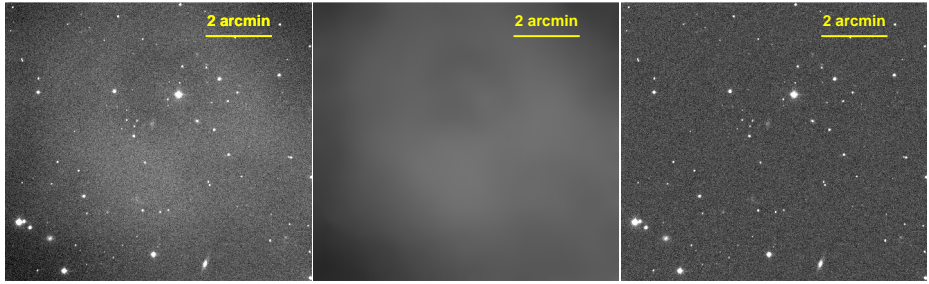}
\caption{Example of the sky background subtraction of LSBG AGC 102672. 
All images size is 9$\arcmin$.0 $\times$ 8$\arcmin$.3, and the length of the yellow line is 2$\arcmin$.
The left panel is the original $R$-band image. The middle panel is the constructed $R$-band sky background, and the right panel is the sky-background-subtracted image. All three images are in the same scale range.}
\label{fig3}
\end{figure*}

\begin{figure*}
\epsscale{0.8}
\plotone{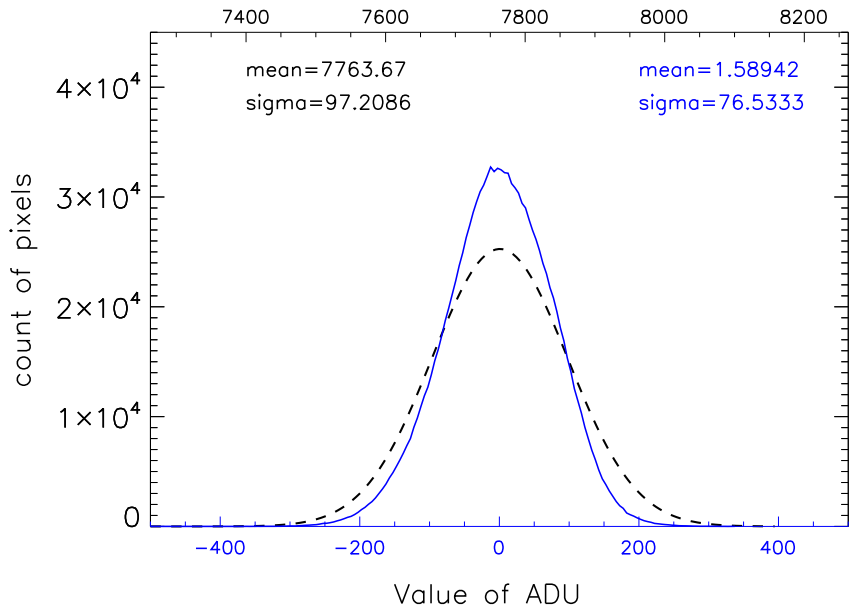}
\caption{Example of distributions of the global background fluctuations of LSBG AGC 102672 before (black dashed line) and after (blue solid line) background subtraction. A gaussian fitting is applied to the two distributions. The upper portion of the panel gives the mean values and standard deviations of the two distributions.}
\label{fig4}
\end{figure*}

The next step is sky background construction, which is the most critical step of data reduction. 
Sextractor is employed to detect faint or extended objects in the gaussian smoothed image.
A mask image is produced after taking all the detected objects off.
In order to obtain the large-scale structures of the background, a median filter of 70$\times$70 $pixel^{2}$ is applied to the mask image to reduce the random noise and to fill in the mask regions with surrounding sky regions.
The constructed sky background image is subtracted from each image.
Figure \ref{fig3} shows an example of the original image, the constructed sky background, and the background-substracted images; all three images are in the same value scale range.
We can see that the sky background reflects the vignetting and non-uniformity distribution.
We also compare the fluctuation of the original and sky-subtracted images in Figure \ref{fig4}.
From Figure \ref{fig4}, the median distribution of the background after being sky-subtracted is more closer to 0.
The fluctuation of sky-subtracted image(blue solid line) is much less than that of the original image (black dashed line).

Since the H$\alpha$ images contain contributions from both H$\alpha$ emission and the underlying stellar continuum, 
it is also important to remove the stellar continuum to get the real H$\alpha$ emission.
 Here, we adopt the $R$-band image as the stellar continuum, due to the fact that the wavelength coverage of $R$-band is wide enough to be dominated by the stellar continuum.
In order to subtract the continuum from the observed H$\alpha$ frames, we must scale the continuum flux of H$\alpha$ to same level as the flux of $R$-band image.
In this process, we assume that field stars have no H$\alpha$ emission, which means that they should have the same continuum flux ratios between H$\alpha$ and $R$-band images.
We define the count ratio of the wide $R$-band and narrow H$\alpha$ band as WNCR:
\begin{equation}
WNCR = \frac{c_{W,cont}}{c_{N,cont}}
\label{equation1}
\end{equation}
Here $c_{W,cont}$ and $c_{N,cont}$ are the measured count of the wide $R$-band and narrow $H\alpha$ band filters. 

Statistically, the median WNCR of these field stars could be treated as the scale factor to subtract the continuum from the H$\alpha$  image.
To obtain an accurate WNCR, we adopt aperture photometry, with the radii of 5 times the FWHM ofthe point-spread-function for stars in each image, using Sextractor and selected field stars with S$/$N greater than 20.
To match the continuum, $\rm H\alpha$ image multiply WNCR and subtract the $R$-band image.
It is tricky to adjust the value around WNCR to get the best scaled one.
Finally the continuum is removed from scaled H$\alpha$ images, when the residual fluxes of most selected field stars reached a minimum.

The scaled values we used are from field stars, However, the scaled value of the object galaxies is somewhat different.
The color of the studied galaxy is different from that of the field stars. 
The color effect of field star would cause errors, leading to underestimates as large as 40$\%$ and overestimates as large as 10$\%$  when measuring $\rm H\alpha$ equivalent width \citep{2012MNRAS.419.2156S}.
To quantify the errors, we selected different spectral types (F,G,K) of stars taken from the MILES stellar library.
 
Because all our sample galaxies are located at high galactic latitude (82 $\%$  sample $> 30\degr$ ) and M stars are too faint, only F G K stars were considered, the WNCR error can be under-estimated as large as 7$\%$ and overestimated as large as 7$\%$.

\begin{figure*}
\epsscale{0.8}
\plotone{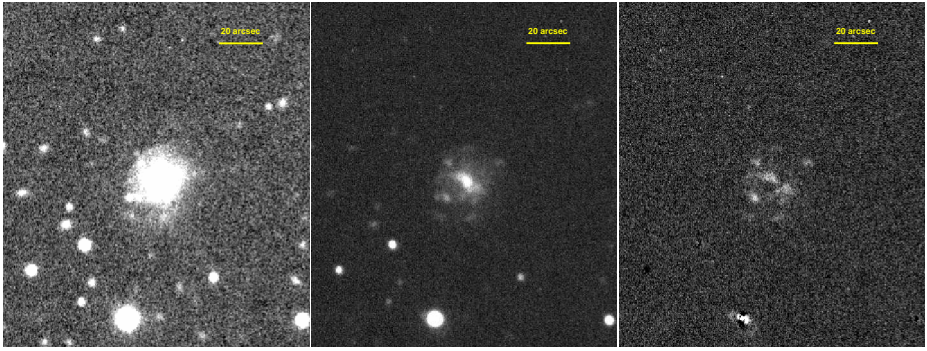}
\caption{ Example images of LSBG AGC 102243 showing the process of continuum subtraction. $R$-band, H$\alpha$ band, and continuum-subtracted H$\alpha$ images are shown in this figure from left to right.}
\label{fig5} 
\end{figure*}

Figure \ref{fig5} shows the $R$-band, H$\alpha$ narrow band and continuum-subtracted H$\alpha$ images of LSBG AGC 102243 from left to right, as an example.
 As Du2015 derived from the SDSS survey, the flux calibrations for the observed broad and narrow band images are undertaken depending on the SDSS photometry. 
The field stars with $S/N > 20$ in both SDSS and our $R$-band image are selected for flux calibration. 
Here, the aperture magnitudes of the SDSS $r$-band and $i$-band are used to calculate the Johnson $R$-band magnitudes based on the Equation \ref{eq:2} \citep{2005AAS...20713308L} as follows. 
The Johnson $R$-band magnitude is transformed to $AB$ magnitude systems with Equation \ref{eq3}  \citep{1994AJ....108.1476F}. Then, the $AB$ magnitude is transformed to flux density with Equation \ref{eq4}.

\begin{equation}
R = r - 0.2936(r - i)-1.439\ ; \sigma = 0.0072 \label{eq:2}
\end{equation}
\begin{equation}
R(AB) = R + 0.055\label{eq3}
\end{equation}
\begin{equation}
m_{AB} = -2.5\ log_{10}(\frac{f_{\nu}}{3631\ JY})\label{eq4}
\end{equation}

Based on this formula, we derive the averaged calibration factor (flux density per count) of each image, which is then applied to calibrating the photometric fluxes in both $R$-band and continuum-subtracted H$\alpha$ images.

\subsection{Photometry}

An elliptical aperture is adopted to perform photometry on both $R$-band and H$\alpha$ band images.
Firstly, the broad $R$-band image is used to determine photometric radius.
Helped by the IRAF task ellipse, we can obtain the profile of the total flux counts enclosed by an elliptical aperture, along with the semimajor axis. 
Then, the flux at which the growth curve reaches 25 $\rm magarcsec^{-2}$, the semimajor axis($a$) and semimini axis($b$) are adopted as the optical photometry radius.
H$\alpha$ flux is total flux enclosed by elliptical area.
There are 111 objects in total  and 19 objects cannot be detected because of their weak $\mathrm{H\alpha}$ emission. 

\section{H$\alpha$ Flux of LSBGs}

\subsection{Flux Correction }

Taking the H$\alpha$ filter transmission cruve into account, we adopt the transmission curve of $H\alpha$ filters in Figure \ref{fig2} and correct the transmission loss brought by the H$\alpha$ filters.
The normalized transmission $\rm T(H\alpha)$  used for correcting the flux is derived from the  equation,
 
\begin{equation}
T(H\alpha) =\frac{ T'(H\alpha)}{  \int_{\lambda1}^{\lambda2}{T'(\lambda)d\lambda}/ FWHM}
\end{equation}

where $T'(\lambda)$ is the transmission curve,
$\rm T'(H\alpha)$ is the direct transmission at galaxy-redshifted H$\alpha$ wavelength from the transmission curve,   
$T(H\alpha)$ is the normalized transmission at the galaxy-redshifted H$\alpha$ wavelength,
and $\lambda1$ and $\lambda2$ are the starting and ending wavelengths of the transmission curve.
FWHM is the full width at half maximum of the H$\alpha$ filters. 
The corrected H$\alpha$ flux is obtained after dividing the normalized transmission $\rm T(H\alpha)$.

The bandwidth of $R$-band filter we used is wide enough, which leads to the fact that, apart from the stellar continuum, the observed flux in the $R$ filter still contains the contribution from H$\alpha$ emission, which will result in the loss of H$\alpha$ flux during the process of stellar continuum subtraction. 
Fortunately, such a loss can be estimated (about 4\%) and corrected according to the bandwidth of both the $R$ and H$\alpha$ filters.

 The extinctions for the galaxies in our sample include the contributions from both Galactic and intrinsic extinctions.
For nearby galaxies, their H$\alpha$ emission feature is covered by the SDSS $r$ filter. 
Therefore, we adopt the extinction value in the SDSS $r$-band to correct observed H$\alpha$ Galactic extinction.
Generally, intrinsic extinction correction is derived from the Balmer emission line ratio of $\rm F_{ H\alpha}/F_{H\beta}$. 
The color excess E(B-V) can be derived from [$\rm F_{H\alpha}/F_{H\beta}$]/[$F_{H\alpha 0}/F_{H\beta 0}$] according to CCM extinction law \citep{1989ApJ...345..245C}. 
Here, we adopt the intrinsic ratio $\rm F_{H\alpha 0}/F_{H\beta 0}$ as 2.87 for \ion{H}{2} galaxies, then the extinction correction of H$\alpha$ flux is calculated from $\rm A_{H\alpha}$ = 2.468E(B-V) \citep{2001PASP..113.1449C}.
However, only 20$\%$ of the LSBGs in our fall sample have nuclear fiber spectra from SDSS. 
Therefore, we have to adopt the same extinction correction and assume that there is no extinction gradient for all sample LSBGs. 
In total, 510 LSBGs from Du2015 have available SDSS spectra  and Balmer ratio $\rm F_{H\alpha}/F_{H\beta}$ derived from the MPA-JHU catalog of SDSS DR7.
Finally, we adopt a median value of $\rm F_{H\alpha}/F_{H\beta}=3.1493$ for the 510 LSBGs as the extinction correction for all sample LSBGs.

Owing to the approximate 60$\AA$ FWHM bandwidth of those H$\alpha$ filters, [\ion{N}{2}]$\lambda\lambda$6548, 6584 features also contribute to the obtained H$\alpha$ images.
We can remove these [\ion{N}{2}] features following equation \ref{eq6} with the assumption of the a fixed ratio of [\ion{N}{2}]/H$\alpha$ throughout all the galaxies.
\begin{equation}
f_{H\alpha,corr[NII]}=\frac{f_{H\alpha+N[II]}}{1+\frac{f_{NII}}{f_{H\alpha}}}\label{eq6}
\end{equation}
Similar to intrinsic extinction correction, we take the median ratio 0.1578 of [\ion{N}{2}]/H$\alpha$ for all 510 LSBGs with available SDSS fiber spectra, and apply it to [\ion{N}{2}] correction for all the galaxies in our sample.

\subsection{ H$\alpha$ Flux and Reliability}
\begin{figure*}
\epsscale{0.8}
\plotone{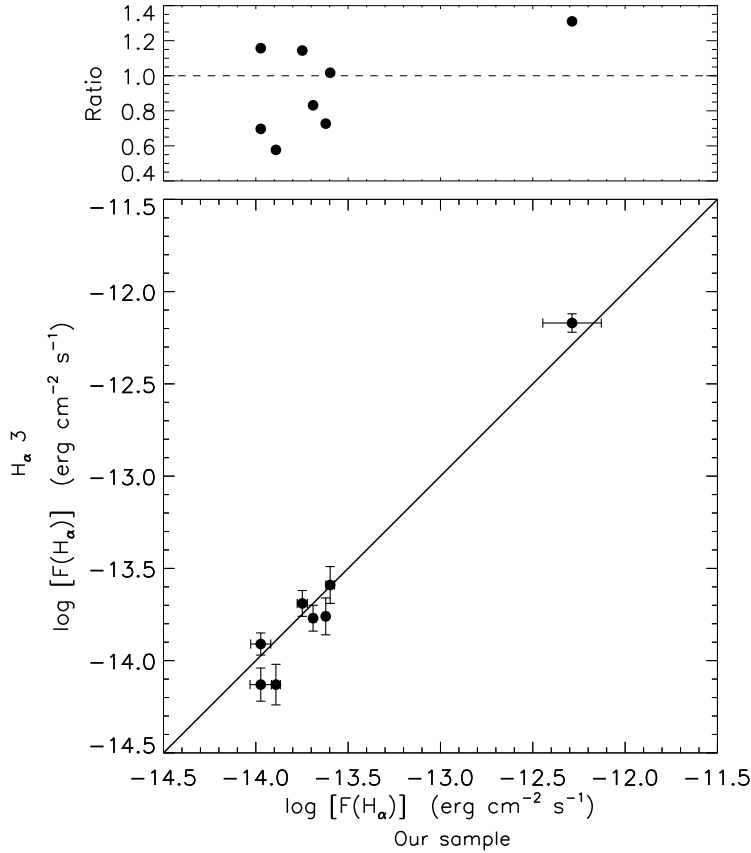}
\caption{ Comparison of the H$\alpha$ flux of eight LSBGs between our measurements and the H$\alpha$3 survey.
          The upper panel is the ratio value ($Flux_{H\alpha 3}/Flux_{our sample}$) between H$\alpha 3$ flux and our sample. }
\label{fig6}
\end{figure*}

After all the corrections above, we get the total H$\alpha$ flux for each LSBG.
In order to compare with previous works, we check eight LSBGs from our spring sample which that also belong to the H$\alpha$3 survey \citep{2015A&A...576A..16G}.
Figure \ref{fig6} shows a comparison between the LSBGs fluxes estimated by us and those derived from the H$\alpha$3 survey, 
and the upper panel is the ratio between H$\alpha$3 survey flux and ours.
The differences between them are around 0.1 dex and less than 0.18 dex.
Roughly speaking, these two calibrated H$\alpha$ fluxes are consistent.

\begin{figure*}
\epsscale{0.8}
\plotone{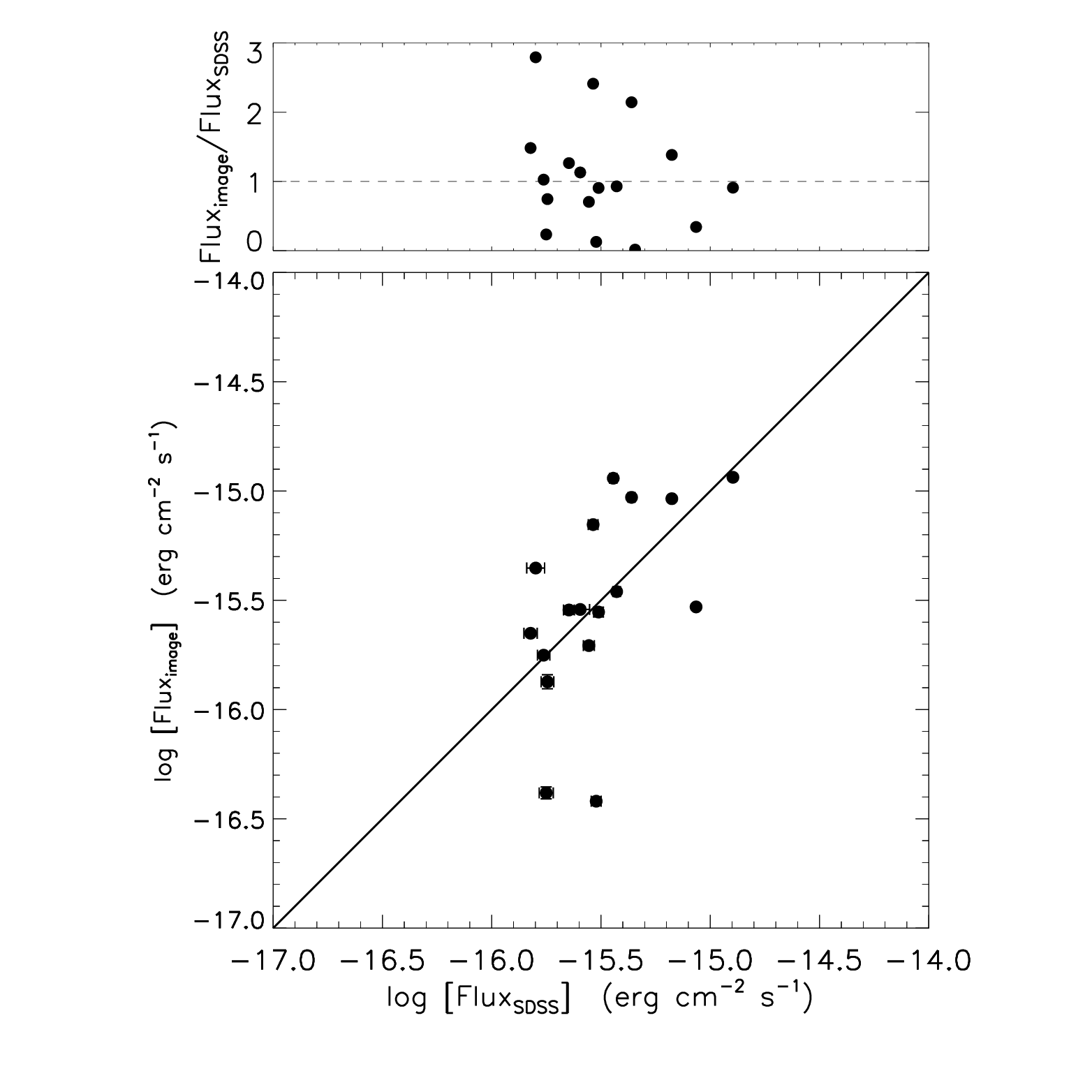}
\caption{ Comparison of our sample flux from our H$\alpha$ image and SDSS fiber spectra flux within 3$\arcsec$. 
         The upper panel shows the flux ratio between the H$\alpha$ image and SDSS spectra. }
\label{fig7}
\end{figure*}

Since 20$\%$ of the LSBGs in our fall sample have SDSS fiber spectra, the H$\alpha$ flux can also be derived directly from the MPA-JHU directly. 
We firstly measure the $H\alpha$ flux on the image within the SDSS fiber diameter(3$\arcsec$) and then compare with H$\alpha$ flux from SDSS fiber spectra in Figure \ref{fig7}.
Most of the H$\alpha$ flux is consistent.
There are two objects that deviate far away from the SDSS fiber flux. 
 After checking with an $H\alpha$ image we found that there is no detectable $H\alpha$ emission where the fiber is located. 

We also check the SFR of these LSBGs.
Due to the 3$\arcsec$ fiber diameter, an aperture correction is needed to get the total H$\alpha$ flux of the whole galaxy.
Here, we assume that the H$\alpha$ emission follows the same distribution as the SDSS $r$-band image. The value of the aperture correction can 
be calculated from the difference between the fiber and Petrosian magnitudes in $r$-band as follows:
\begin{equation}
F_{Petro}/F_{Fiber}=10^{-0.4(m_{petro}-m_{fiber})}.
\end{equation}
Here, $\rm m_{petro}$ and $\rm m_{fiber}$ are Petrosian and fiber magnitudes in the $r$-band, respectively.
$\rm F_{Fiber}$ represents the H$\alpha$ flux of a galaxy in the given fiber aperture,
whereas $F_{petro}$ is the total H$\alpha$ flux inside the Petrosian aperture.

\begin{figure*}
\epsscale{0.8}
\plotone{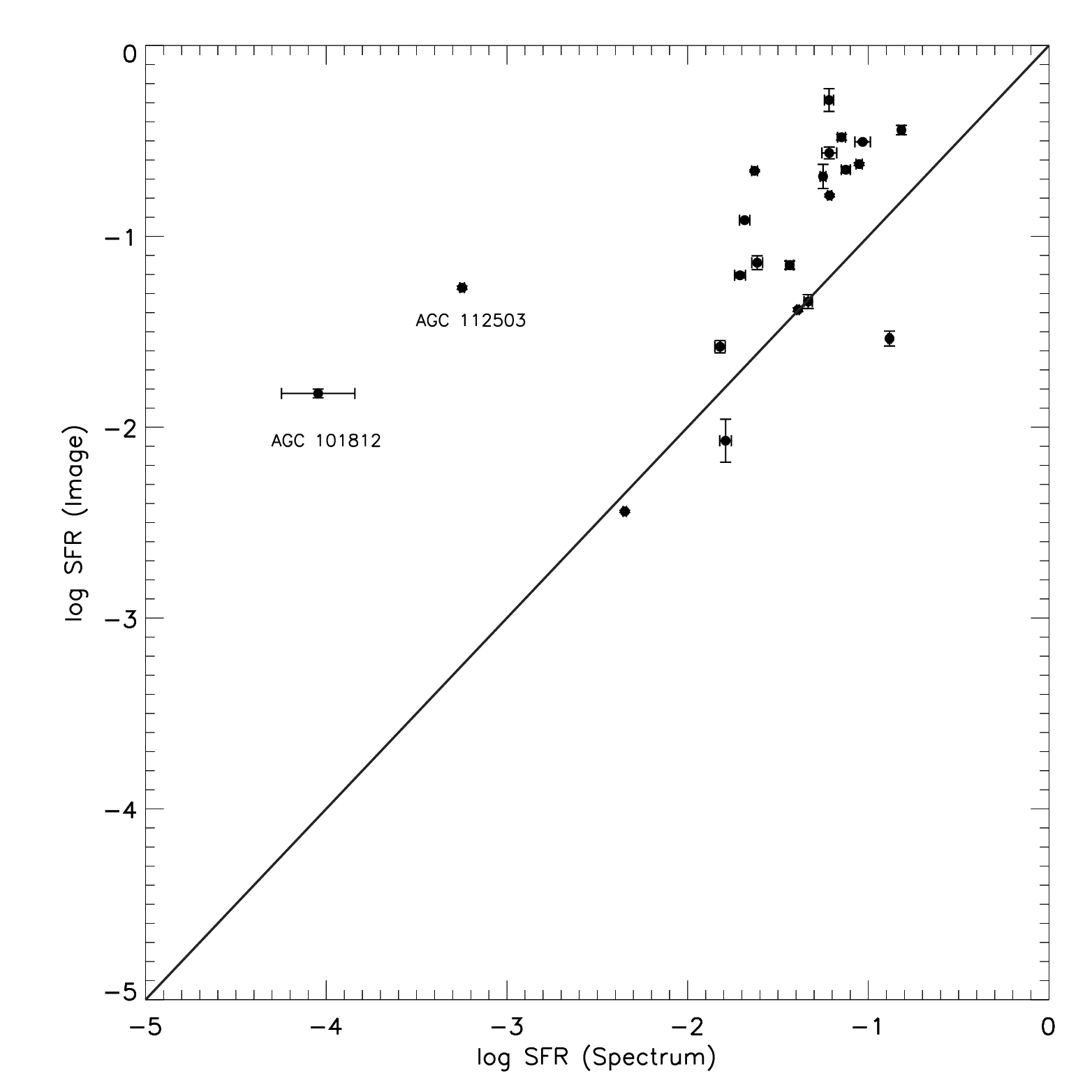}
\caption{ Comparison of the measured SFRs of 22 LSBGs derived by aperture-corrected SDSS fiber spectra.}
\label{fig8}
\end{figure*}

H$\alpha$ emission traces the location of the star formation region and also provides a fairly robust quantitative measure of its current SFR. 
The SFR of the LSBGs in our sample is calculated from the H$\alpha$ luminosity and using the following calibration \citep{1998ApJ...498..541K}.
\begin{equation}
SFR_{H\alpha}(M_{\sun}\ yr^{-1}) = 7.9 \times 10^{-42}[L(H\alpha)](erg\ s^{-1})
\end{equation}
where L(H$\alpha$) is the intrinsic extinction-corrected H$\alpha$ luminosity.
The initial mass function used in the conversion is a Salpeter function $[dN(m)/dm=-2.35]$ over $m=0.1-100 M_{\sun}$.
Figure \ref{fig8} shows the a comparison between the SFRs of LSBGs calculated from an H$\alpha$ image and H$\alpha$ spectrum. 
For most of the LSBGs in our sample, the SFRs derived from SDSS spectra are less than those from H$\alpha$ images, and there are two LSBGs (AGC 101812, AGC 112503) showing large deviations,  probably due to the aperture correction.
Checking the SDSS images of AGC 101812 and AGC 112503 shows that there exist several bright blue knots outside of the fiber region.
Thus, aperture corrections have largely underestimated the total H$\alpha$ emission.
Therefore, it isinadequate to calculate the total H$\alpha$ flux for the entire galaxy solely  from the fiber spectrum.

All the H$\alpha$ flux and other basic parameters of LSBGs are listed in Table \ref{table3}.
The table columns can be briefly described as follows:\\
Column 1 $:$ galaxy name in terms of AGC number.\\
 Column 2$:$ the semimajor axis from elliptical photometry (kpc), which is the radii at 25 $magarcsec^{-2}$. \\
Column 3 $:$ the ellipticity from the elliptical photometry.\\
Column 4 and 5 $:$ logarithm of the  H$\alpha$ flux and  error ($erg\ s^{-1} cm^{-2}$).\\
Column 6 $:$ the logarithm of the SFR ($M_{\sun} yr^{-1}$).\\
Column 7 $:$ the logarithm of the SFR surface density ($M_{\sun} yr^{-1} kpc^{-2}$).\\
Column 8 $:$ the logarithm of the {\hi} mass taken from the $\alpha.$40 catalog \citep{2011AJ....142..170H}.\\
Column 9 $:$ the logarithm of the {\hi} gas surface density ($M_{\sun} pc^{-2}$).

We will explore the SFR and SFR surface density, and {\hi} gas and {\hi} gas surface density in the next section.

\section{Results and Analysis }


\subsection{The Star Formation and Gas Surface Density}

For each LSBG in our sample, the enclosed region of elliptical photometry is used as the optical area to calculate the star formation surface density($\Sigma_{SFR}$).
For the majority of the targets, the beam size of ALFALFA {\hi} observation is 3.5 arcmin, which is too large to obtain a suitable {\hi} size.
Hence, we have to derive the {\hi} size from the calibrated optical photometry size. 
$r_{HI}/r_{25}$ is almost constant (1.7$\pm$0.5) and shows weak dependence on the type from S0 to Im  \citep{1997A&A...324..877B,2002A&A...390..863S,2015ApJ...808...66J}. We adopt 1.7 times the optical photometry radii as the {\hi} radii.
Hence, the {\hi} surface density $\Sigma_{HI}$ is calculated from the following Equation:
\begin{equation}
\Sigma_{HI}=\frac{M_{HI}}{\pi\ (1.7^{2}ab)}
\end{equation}
Here, $M_{HI}$ is the {\hi} mass derived from the ALFALFA catalog, and a and b are the semimajor and semi-minor radii of photometry ellipticals, respectively.
SFE is defined as the ratio of SFR and gas mass. Generally, the gas in a normal galaxy consists of ionized, atomic, and molecular gas. 
Since our sample is an {\hi}-selected sample and lacks of molecular observations, we just calculate $SFE_{HI}$ as follows: 
\begin{equation}
  SFE_{HI}= \frac{SFR}{M_{HI}}
\end{equation}

\begin{figure*}
\epsscale{1.0}
\plotone{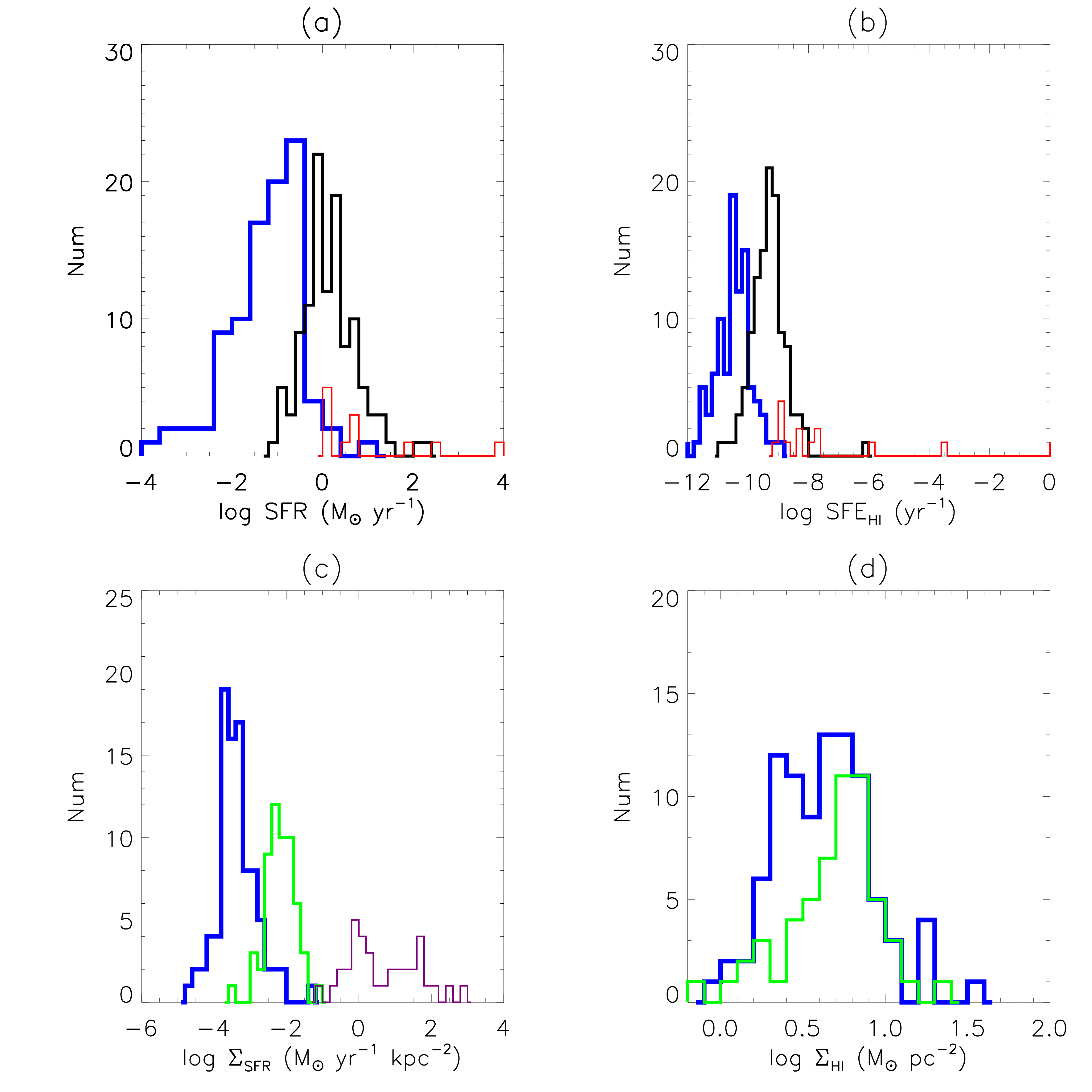}
\caption{ Distributions of (a) star formation rate; (b) star formation efficiency, SFE=SFR/mass({\hi}); (c) star formation surface density; (d) gas ({\hi}) surface density.
Blue represents the LSBGs in this paper.
 The black and red  colors in (a) and (b) represent star forming galaxies from \citet{1996AJ....112.1903Y} and starburst galaxies from \citet{2015ApJ...808...66J}.
 The green (c) and purple (d)  represent star forming galaxies and starburst galaxies from \citet{1998ApJ...498..541K}, respectively.}
\label{fig9}
\end{figure*}

The distributions of the SFR, $SFE_{HI}$ and $\Sigma_{SFR}$, $\Sigma_{HI}$ are shown in Figure \ref{fig9}.
For comparison, we also show the distributions for samples of star forming and starburst galaxies.
In panels (a) and (b), star-forming galaxies are derived from \citet{1996AJ....112.1903Y}, and star burst galaxies are from \citet{2015ApJ...808...66J}.
In panels (c) and (d), both star forming galaxies and starburst galaxies are derived from \citet{1998ApJ...498..541K}.
Compared with star-forming and starburst galaxies, both the SFR and $SFE_{HI}$ of LSBGs are lower than those of star forming galaxies by approximately one order of magnitude, and even far lower than those of starburst galaxies. Furthermore, the SFR surface densities $\Sigma_{SFR}$ of LSBGs are even about more than one order of magnitudes lower than those of star forming galaxies.
\subsection{Kennicutt-Schmidt Law}

\begin{figure*}
\epsscale{0.9}
\plotone{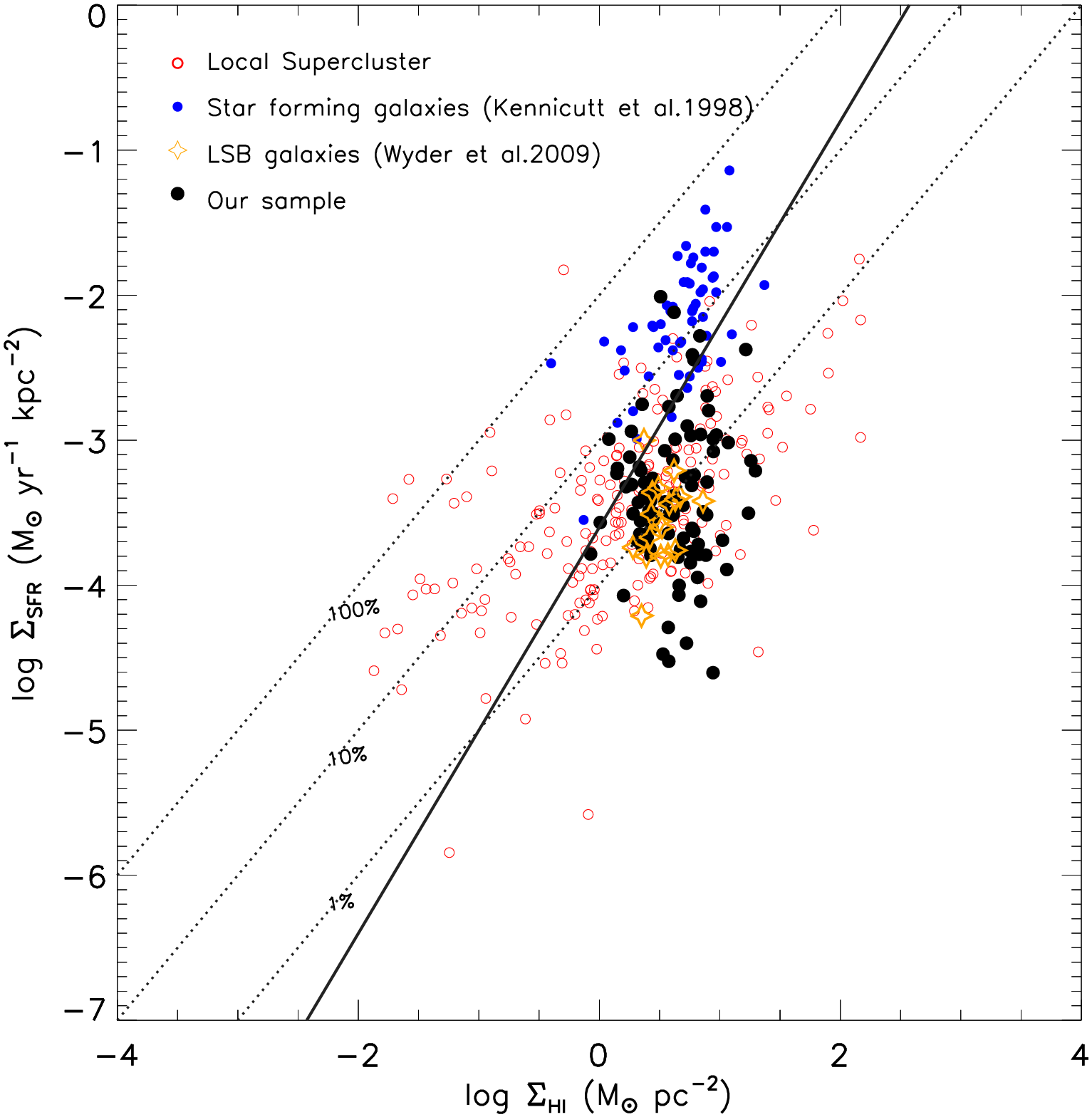}
\caption{Relation between SFR surface density and {\hi} surface density.
 The black dots are from this paper, the yellow diamonds are LSBGs from \citet{2009ApJ...696.1834W}, and the blue dots are star forming galaxies from \citet{1998ApJ...498..541K}. 
The red circles are galaxies in the Local Supercluster in the $H\alpha$3 survey from  \citet{2012A&A...545A..16G}.
The black solid line is the Kennicutt-Schmidt Law, and the three dotted lines show the {\hi} SFEs of 100\%,10\%,1\% in a timescale of star formation of $10^{8}$ yr.} 
\label{fig10}
\end{figure*}

Figure \ref{fig10} shows the relation between SFR surface density and {\hi} surface density ($\Sigma_{HI}$) .   
The blue symbols are star forming (disk) galaxies from \citet {1998ApJ...498..541K}.
The red circles are galaxies belonging to the Local supercluster \citep{2012A&A...545A..16G} and the black points are LSBGs in our sample.
The orange stars are LSBGs from \citet{2009ApJ...696.1834W}.
Following \citet{2003ApJ...588..230O}, we plot dotted lines with SFEs of $1\%$, $10\%$, and $100\%$ in a timescale of star formation of $10^{8}$ yr, corresponding to typical orbital timescales in galaxies.
The Kennicutt-Schmidt law is plotted as a black solid line.
The coverage of our LSBGs is similar to that of \citet{2009ApJ...696.1834W} LSBGs, but is toward to even lower star formation surface density.
From Figure \ref{fig10}, LSBGs and star forming galaxies are in the same region of the {\hi} surface density, but LSBGs have much lower SFR surface densities than star-forming galaxies.
Galaxies in the Local Supercluster have a more diffuse $\Sigma_{HI}$ distribution.  

Several previous works tried to detect CO emission in LSBGs.
However, most of them only gave upper limits on CO content, and a few LSBGs detected molecular gas.\citep{2001ApJ...549L.191M,2003ApJ...588..230O,2005AJ....129.1849M,2010A&A...523A..63D}.
\citet{2017arXiv170801362C} observed the CO (2-1) of nine LSBGs from Du2015 with JCMT, 
but none of them is detected CO (2-1) emission, so only upper limits $M_{H_{2}}$ are given.
The $M_{H_{2}}/M_{HI}$ ratios are less than 0.02, which indicates a shortage of molecular gas in LSBGs \citep{2017arXiv170801362C}.

\begin{figure*}
\epsscale{0.9}
\plotone{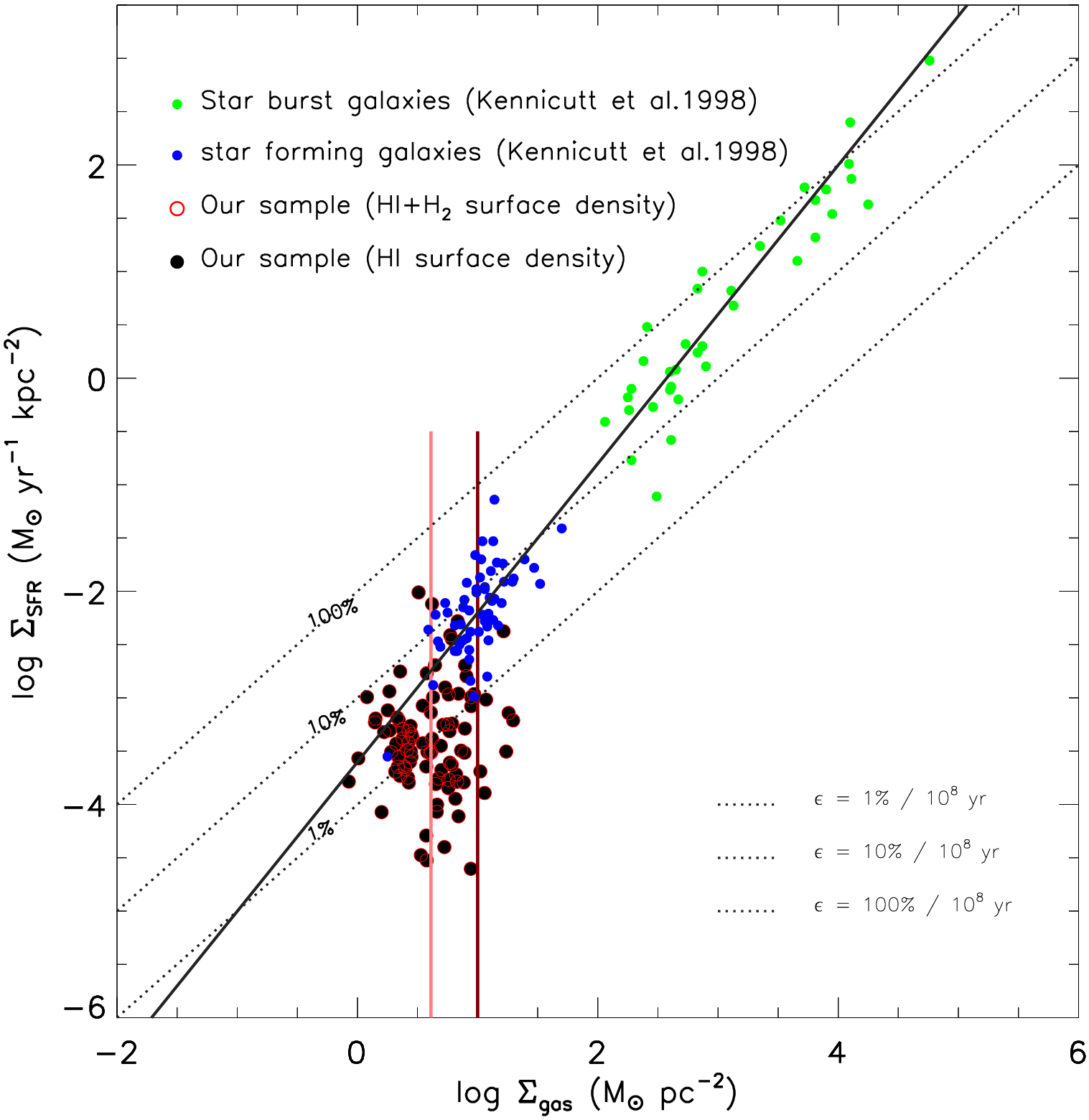}
\caption{ Relation between SFR surface density and {\hi} surface density.
          our LSBG sample is the black solid circle ({\hi} gas surface density) and red circles (gas surface density).
         The blue dots are star forming galaxies and the green dots are starburst galaxies; both are from \citet{1998ApJ...498..541K}.
        The black solid line is the Kennicutt-Schmidt Law, and the three dotted lines show the {\hi} SFEs of 100\%,10\%,1\% on a timescale of star formation of $10^{8}$ yr. 
       The pink line is the mean value of the LSBG gas surface density and the brown dashed line is the upper boundary of the low gas surface density of 10 $ M_{\odot}pc^{-2}$.
}
      \label{fig11}
\end{figure*}

\citet{2008AJ....136.2846B} derived a correlation between SFR surface brightness density and $\rm H_{2}$ surface density,
\begin{equation}
 \Sigma_{SFR} =10^{-2.1\pm0.2} \Sigma_{H2}^{1.0\pm0.2}
\end{equation}
which helps us to estimate the approximate $\rm H_{2}$ surface density from this relation.
 Even though $H_{2}$ gas is not distributed as the {\hi} gas \citep{2008AJ....136.2782L,2011A&A...534A.102L}, Equation \ref{fig11} can be used as a rough estimation of $\Sigma_{H_{2}}$.
To get accurate values, future interferometric  {\hi} and CO data are necessary.

From Figure \ref{fig11}, gas surface density $\Sigma_{HI+H_{2}}$ (red circles) is very close to  $\Sigma_{HI}$ (black dots), which is consistent with our previous assumption: {\hi} dominates the gas content of our LSBGs.

All LSBGs are located at the cutoff region, deviating from the kennicutt-Schmidt law (black line), which is derived from the star-forming (blue dots) and starburst galaxies (green dots). 

 According to the dashed line (SFE), starburst galaxies have SFEs that are higher than 10$\%$, and star forming galaxies have SFEs a little lower than 10\%, but still much higher than 1$\%$. Though a small number of LSBGs are blended with star forming galaxies, LSBGs have SFEs far below those  of star forming galaxies and of around 1\% for most of them. In some extreme cases, SFE can even be lower than 0.1$\%$.
There is a special LSBG, AGC 748765,  whose SFE is far above 10$\%$. It has an extremely luminous \ion{H}{2} region in its disk.

\citet{2012ARA&A..50..531K} pointed out that the gas surface density can crudely be divided into three regions$:$ low density ($\Sigma_{gas}<10 M_{\odot}pc^{-2}$), intermediate density($10 M_{\odot}pc^{-2} < \Sigma_{gas}< 100-300M_{\odot}pc^{-2}$), and high density ($\Sigma_{gas}>100-300 M_{\odot}pc^{-2}$). 
Although the SFR surface density of LSBGs can spread more than three orders of magnitudes, their gas surface densities are in a narrow region within one order of magnitude from 1 to 10  $M_{\odot}pc^{-2}$. 
The SFR surface density of LSBGs does not show any dependence on gas or {\hi} surface density.
The brown line is the upper limit for the low-density region in Figure \ref{fig11}.
The mean gas surface density for LSBGs in our sample ($\Sigma_{gas}= 4.1 M_{\odot}pc^{-2}$) is shown as a pink line in Figure \ref{fig11}.
As expected, LSBGs are located in the low-density region. However, many star forming galaxies are also located in the low-density region, but with higher SFR density.
The tight relation between SFR and molecular gas \citep{2004ApJS..152...63G,2008AJ....136.2846B} demonstrates that the molecular gas could still dominate the gas in star forming galaxies. From Figure \ref{fig11}, the turnff point of the K-S Law is at around $\Sigma_{gas}=4 M_{\odot}pc^{-2}$, which is almost the lowest gas density of star forming galaxies, and also a the similar value to the mean gas surface density of LSBGs.
What causes that the SFR surface density to be widely distributed in such low-density regions is worth exploring in the future work.

\subsection{Star Formation History }

\begin{figure*}
\epsscale{0.9}
\plotone{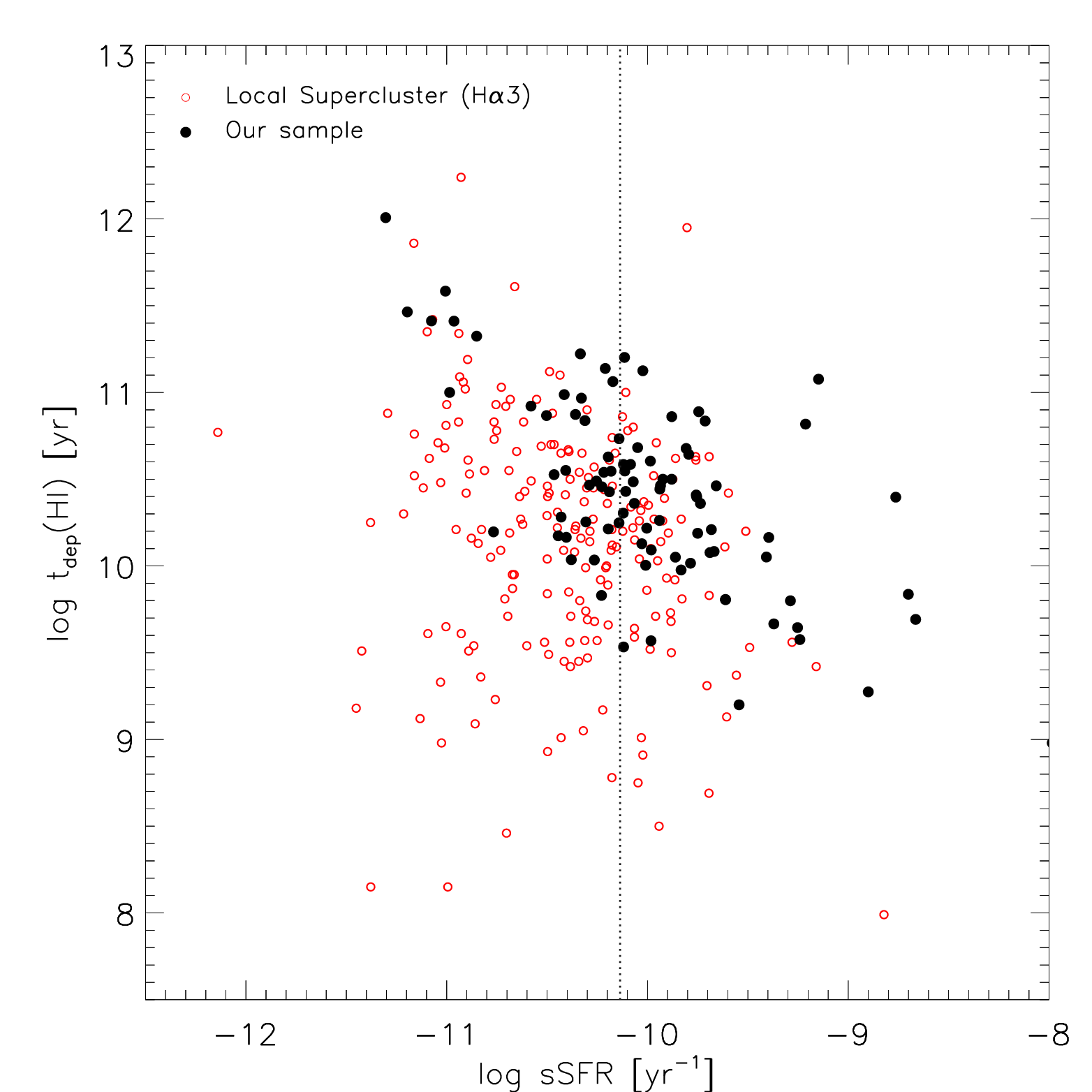}
\caption{
$t_{dep}$({\hi}) vs. sSFR diagram.  The black solid circles are our LSBG sample, and the red open circles are galaxies in the Local Supercluster from the $H\alpha$3 survey \citep{2012A&A...545A..16G}.
}
\label{fig12}
\end{figure*}

To characterize the evolutionary status of the star formation in galaxies, we follow specific ($sSFR=SFR/M_{*}$) and HI depletion time($t_{dep}(HI)=M_{HI}/SFR$) to study the star formation history of LSBGs.
Stellar mass is derived from $g$- and $r$-band magnitudes from Du2015 follows the equation $log(M_{*}/M_{\sun})=-0.306+1.097*(g-r)+log{L_{r}/L_{\sun}}$ \citep{2003ApJS..149..289B}.
{\hi} depletion time and sSFR relation are shown in Figure \ref{fig12}.
The red circles are galaxies from the Local Supercluster\citep{2012A&A...545A..16G} and the black solid circles are our LSBGs.
The dashed line representing the sSFR value is -10.1367, which means a galaxy can gain current stellar mass in current SFR throughout the Hubble time.
Here, Hubble time is adopted with 13.7 Gyr \citep{2007ApJS..170..377S}.
The dashed line is the boundary between the active phase of galaxies and the quiescent phase.

On average, the current SFRs in the Local Supercluster cannot account for their current masses, though they present higher SFRs than those of LSBGs.
Galaxies in the Local Supercluster should experience intensive star formation events once or several times in their star formation histories.
Most LSBGs are around the dashed line and some LSBGs are active phase galaxies.
Even with such a low current SFRs, most LSBGs can still obtain the current stellar mass over the timescale of universe. 
They do not need  a strong interacting or major merging process to occur. 
A stochastic and sporadic star formation scenario could explain such a low and stable star formation histories \citep{1995MNRAS.274..235D,2015MNRAS.446.4291L}.
 The lower number density environment of LSBGs may indicate that they seldom experience galactic interacting or merging \citep{2015AJ....149..199D}. This is supported by the stellar populations with ages around 2 Gyr in LSBGs \citep{2017ApJ...837..152D}.
The higher $t_{dep}(HI)$ of our LSBGs suggest that they will have an abundant supply of {\hi} in the future.

\section{SUMMARY}

We performed a narrow band H$\alpha$ imaging survey for LSBGs selected from the 40\% ALFALFA extragalactic {\hi} survey. 
A sample of 111 LSBGs in the fall sky has been observed with the Xinglong 2.16m telescope. 
The LSBGs in this sample have recession velocities ranging from 1012 to 9889 Km$\thinspace$$s^{-1}$ and {\hi} masses from $log_{10}M_{HI}=7.73$ to $log_{10}M_{HI}=10.14$. H$\alpha$ fluxes of 92 objects are measured using IRAF ellipse photometry.
The derived total H$\alpha$ fluxes and corresponding SFRs are listed in Table \ref{table3}.
All the LSBGs in our sample have blue features that are similar to those of other LSBG samples. 
They have lower SFRs, lower SFEs, lower star formation surface densities, lower gas surface densities and similar {\hi} surface densities compared with normal star forming galaxies.

Most of LSBGs are in low surface density regions and are below the Kennicutt$-$Schmidt relation. 
Their SFR surface densities spread about three orders of magnitude and their SFE efficiencies are around 1$\%$ or even lower.
To characterize the star formation histories of LSGBs, we adopt parameters of $t_{dep}$({\hi}) and sSFR. 
From the distribution of both parameters, LSBGs tend to be gas-rich and their star formation histories tend to be stable, rarely suffering from intensive interaction or major mergers.

\section*{Acknowledgments}

We thank the referee for constructive comments and suggestions. This project is supported by the National Key R$\&$D Program of China (No.2017YFA0402704), and the National Natural Science Foundation of China (grant No. 11733006,11403037, 11225316, 11173030, 11303038,  11403061 and U1531245), the China Ministry of Science and Technology under the State Key Development Program for Basic Research (2014CB845705), and the Strategic Priority Research Program, The Emergence of Cosmological Structures of the Chinese Academy of Sciences (grant No.XDB09000000).
This project is also supported by the China Ministry of Science and Technology under the State Key
Development Program for Basic Research (grant No.2014CB845705).

We acknowledge the support of the staff of the Xinglong 2.16m telescope. 
This work is partially supported by the Open Project Program of the Key Laboratory of Optical Astronomy, National Astronomical Observatories, Chinese Academy of Sciences.
We also thank the ALFALFA team and the SDSS team for the released data.
The Arecibo Observatory is part of the National Astronomy and Ionosphere Center, which is operated by Cornell University under a cooperative agreement with the National Science Foundation. 
The authors acknowledge the work of the entire ALFALFA collaboration team in observing, flagging, and extracting the catalog of galaxies used in this work. The ALFALFA team at Cornell is supported by NSF grant AST-0607007 and AST-1107390 and by grants from the Brinson Foundation.
The authors are thankfull for the useful SDSS database and the MPA/JHU catalogs.
Funding for the SDSS  has been provided by the Alfred P. Sloan Foundation, the Participating Institutions,
the National Science Foundation, the U.S. Department of Energy, the National Aeronautics and Space Administration, the Japanese Monbukagakusho, the Max Planck Society, and the Higher Education Funding Council for England.

\clearpage

\startlongtable
\begin{deluxetable*}{rrrrrrrr}
\tablewidth{0pt}
\tablecolumns{8}
\tablecaption{The Observed Sample of LSBGs \label{table2}}
\tablehead { 
\colhead{AGC} & \colhead{$\mu_{0}(B)$} & \colhead{R.A.} & \colhead{Decl.} &   \colhead{$z$} & \colhead{Dist} & \colhead{Filter} &  \colhead{Date} \\
\colhead{}& \colhead{$\rm magarcsec^{-2}$} &    \colhead{J2000} & \colhead{J2000}& \colhead{}&\colhead{Mpc}&\colhead{} & \colhead{} \\
\colhead{(1)} &\colhead{(2)} &\colhead{(3)} &\colhead{(4)} &\colhead{(5)} &\colhead{(6)} &\colhead{(7)} &\colhead{(8)}  }
\startdata
  17     &    23.4405 &    00:03:43 &    +15:13:05 &    0.0029 &     12.8  &      Ha1     & 20140102 \\
  273    &    22.6556 &    00:28:07 &    +25:59:47 &    0.0187 &     78.9  &      Ha4     & 20140819 \\
  337    &    22.5591 &    00:34:25 &    +24:36:13 &    0.0178 &     74.9  &      Ha3     & 20141021 \\
  1084   &    23.6324 &    01:31:22 &    +23:57:14 &    0.0114 &     46.4  &      Ha3     & 20160206 \\
  1211   &    23.1475 &    01:43:55 &    +13:48:22 &    0.0080  &     32.4  &      Ha2     & 20141017 \\
  1362   &    22.9079 &    01:53:51 &    +14:45:50 &    0.0264 &     109.1 &      Ha5     & 20160108 \\
  1693   &    23.2348 &    02:12:03 &    +14:06:14 &    0.0127 &     52.1  &      Ha3     & 20131010 \\
  2144   &    22.7754 &    02:39:30 &    +29:15:35 &    0.0160  &     65.5  &      Ha3     & 20141021 \\
  12289  &    22.6539 &    22:59:41 &    +24:04:29 &    0.0339 &     140.2 &      Ha6     & 20140826 \\
  12845  &    22.7974 &    23:55:42 &    +31:53:59 &    0.0162 &     69.1  &      Ha3     & 20141021 \\
  100037 &    22.9236 &    00:06:03 &    +27:20:54 &    0.0106 &     44.8  &      Ha2     & 20160205 \\
  100350 &    23.9767 &    00:37:44 &    +24:12:28 &    0.0155 &     65.0  &      Ha3     & 20131231 \\
  101191 &    23.2293 &    00:23:39 &    +15:04:03 &    0.0177 &     74.7  &      Ha3     & 20131010 \\
  101812 &    23.6638 &    00:08:49 &    +14:02:01 &    0.0064 &     27.0  &      Ha2     & 20131230 \\
  101877 &    23.0745 &    00:02:15 &    +14:29:16 &    0.0172 &     72.9  &      Ha3     & 20131010 \\
  101942 &    22.9001 &    00:12:29 &    +15:33:22 &    0.0188 &     79.7  &      Ha4     & 20131010 \\
  101986 &    22.8072 &    00:20:49 &    +15:03:13 &    0.0254 &     104.0 &      Ha4     & 20140825 \\
  102098 &    22.6041 &    00:39:04 &    +14:36:01 &    0.0418 &     174.1 &      Ha7     & 20160208 \\
  102101 &    23.0727 &    00:39:25 &    +14:27:23 &    0.0180  &     75.6  &      Ha3     & 20140819 \\
  102229 &    22.5805 &    00:38:24 &    +25:26:10 &    0.0110  &     45.9  &      Ha2     & 20141017 \\
  102243 &    22.5023 &    00:05:05 &    +23:58:11 &    0.0219 &     89.0  &      Ha4     & 20140825 \\
  102302 &    22.9648 &    00:12:48 &    +14:31:31 &    0.0061 &     25.7  &      Ha2     & 20131230 \\
  102558 &    23.0152 &    00:07:05 &    +27:01:28 &    0.0099 &     41.7  &      Ha2     & 20160205 \\
  102630 &    24.6769 &    00:13:13 &    +25:36:14 &    0.0208 &     88.4  &      Ha4     & 20140101 \\
  102635 &    22.5726 &    00:16:12 &    +24:50:57 &    0.0316 &     130.8 &      Ha5     & 20151210 \\
  102672 &    22.5757 &    00:46:24 &    +25:04:14 &    0.0176 &     73.9  &      Ha3     & 20140819 \\
  102674 &    23.3246 &    00:49:14 &    +25:17:35 &    0.0463 &     193.8 &      Ha7     & 20141017 \\
  102684 &    23.9342 &    00:22:07 &    +25:29:09 &    0.0248 &     101.6 &      Ha4     & 20141018 \\
  102728 &    22.8543 &    00:00:21 &    +31:01:19 &    0.0019 &     9.1   &      Ha1     & 20140101 \\
  102729 &    22.5906 &    00:00:32 &    +30:52:09 &    0.0154 &     65.4  &      Ha3     & 20131231 \\
  102730 &    22.6859 &    00:00:39 &    +31:56:18 &    0.0421 &     175.8 &      Ha7     & 20141017 \\
  102900 &    25.0466 &    00:04:39 &    +29:35:56 &    0.0405 &     168.6 &      Ha6     & 20140827 \\
  102981 &    22.5401 &    00:02:56 &    +28:16:38 &    0.0153 &     64.8  &      Ha3     & 20140820 \\
  110150 &    22.7571 &    01:14:45 &    +27:08:06 &    0.0121 &     49.5  &      Ha3     & 20140819 \\
  110319 &    24.6744 &    01:25:17 &    +14:08:55 &    0.0168 &     69.9  &      Ha3     & 20141021 \\
  110379 &    23.5195 &    01:30:15 &    +14:40:39 &    0.0082 &     33.1  &      Ha2     & 20141017 \\
  110398 &    22.5212 &    01:31:46 &    +14:09:20 &    0.0225 &     92.3  &      Ha4     & 20140825 \\
  112503 &    22.5698 &    01:38:00 &    +14:58:58 &    0.0025 &     10.2  &      Ha1     & 20141017 \\
  112892 &    22.5765 &    01:20:16 &    +14:52:29 &    0.0370  &     154.3 &      Ha6     & 20160210 \\
  113200 &    22.7375 &    01:56:19 &    +14:55:29 &    0.0248 &     102.3 &      Ha4     & 20160210 \\
  113752 &    23.1777 &    01:18:06 &    +27:11:17 &    0.0414 &     173.3 &      Ha6     & 20140827 \\
  113790 &    23.2295 &    01:13:02 &    +27:38:13 &    0.0165 &     68.7  &      Ha3     & 20140826 \\
  113825 &    22.8080  &    01:43:27 &    +24:46:47 &   0.0128 &     52.4  &      Ha3     & 20141021 \\
  113845 &    22.7484 &    01:17:22 &    +24:08:16 &    0.0273 &     112.7 &      Ha5     & 20160108 \\
  113907 &    22.6913 &    01:13:56 &    +30:09:25 &    0.0342 &     142.5 &      Ha6     & 20160208 \\
  113918 &    22.8345 &    01:22:59 &    +32:10:44 &    0.0355 &     148.1 &      Ha6     & 20160208 \\
  113923 &    23.2240  &    01:26:13 &    +32:08:11 &   0.0140  &     57.6  &      Ha3     & 20140819 \\
  114040 &    22.6830  &    01:18:27 &    +29:06:55 &   0.0262 &     108.0 &      Ha4     & 20141018 \\
  121174 &    26.3089 &    02:38:16 &    +29:54:23 &    0.0023 &     9.7   &      Ha1     & 20141017 \\
  122138 &    27.1803 &    02:33:16 &    +28:10:44 &    0.0034 &     13.7  &      Ha1     & 20131230 \\
  122210 &    23.3301 &    02:31:33 &    +26:47:49 &    0.0152 &     62.2  &      Ha3     & 20140826 \\
  122211 &    23.9519 &    02:31:37 &    +26:32:32 &    0.0123 &     49.8  &      Ha3     & 20141021 \\
  122341 &    22.8250  &    02:11:29 &    +14:28:04 &   0.0375 &     156.8 &      Ha6     & 20160208 \\
  122874 &    22.6132 &    02:26:15 &    +24:26:02 &    0.0213 &     87.8  &      Ha4     & 20151210 \\
  122877 &    24.1814 &    02:27:32 &    +24:52:12 &    0.0203 &     85.0  &      Ha4     & 20160210 \\
  122884 &    23.1394 &    02:32:53 &    +25:09:11 &    0.0081 &     32.4  &      Ha2     & 20141017 \\
  122924 &    24.0678 &    02:34:43 &    +24:29:12 &    0.0322 &     134.5 &      Ha5     & 20160207 \\
  123046 &    23.0294 &    02:41:12 &    +31:29:29 &    0.0160  &     65.7  &      Ha3     & 20141021 \\
  123047 &    23.0369 &    02:41:48 &    +31:27:26 &    0.0340  &     142.7 &      Ha6     & 20141017 \\
  123170 &    22.8858 &    02:44:03 &    +29:17:17 &    0.0030  &     12.1  &      Ha1     & 20141017 \\
  123172 &    22.5982 &    02:47:23 &    +29:10:32 &    0.0180  &     74.5  &      Ha3     & 20151204 \\
  320466 &    24.4058 &    22:57:22 &    +27:58:50 &    0.0098 &     43.3  &      Ha2     & 20131230 \\
  321166 &    24.0888 &    22:55:49 &    +14:45:15 &    0.0094 &     41.3  &      Ha2     & 20131230 \\
  321341 &    22.7167 &    22:52:16 &    +24:06:09 &    0.0404 &     168.3 &      Ha6     & 20140827 \\
  321348 &    23.0037 &    22:47:44 &    +23:59:59 &    0.0315 &     129.9 &      Ha5     & 20140826 \\
  321385 &    22.5978 &    22:59:15 &    +24:42:34 &    0.0242 &     98.5  &      Ha4     & 20141018 \\
  321429 &    23.6289 &    22:41:27 &    +31:31:48 &    0.0126 &     55.6  &      Ha3     & 20131231 \\
  321435 &    22.6576 &    22:47:44 &    +32:11:18 &    0.0129 &     56.6  &      Ha3     & 20140819 \\
  321438 &    24.0373 &    22:50:17 &    +30:15:08 &    0.0265 &     108.6 &      Ha5     & 20140101 \\
  321451 &    22.5598 &    22:48:03 &    +29:49:48 &    0.0237 &     96.8  &      Ha4     & 20140820 \\
  321490 &    23.2068 &    22:47:45 &    +28:54:26 &    0.0233 &     95.0  &      Ha4     & 20140825 \\
  321492 &    23.1417 &    22:53:23 &    +29:00:52 &    0.0068 &     30.8  &      Ha2     & 20131230 \\
  331052 &    22.9473 &    23:59:45 &    +27:15:14 &    0.0156 &     66.5  &      Ha3     & 20140819 \\
  332431 &    22.8894 &    23:07:46 &    +14:22:34 &    0.0246 &     100.3 &      Ha4     & 20140820 \\
  332640 &    23.0183 &    23:24:43 &    +13:48:36 &    0.0265 &     108.2 &      Ha5     & 20140825 \\
  332761 &    23.0965 &    23:31:11 &    +15:01:58 &    0.0193 &     82.6  &      Ha4     & 20140825 \\
  332786 &    22.5704 &    23:36:09 &    +15:44:38 &    0.0134 &     57.5  &      Ha3     & 20131229 \\
  332844 &    22.8276 &    23:51:24 &    +14:14:02 &    0.0394 &     163.5 &      Ha6     & 20141021 \\
  332861 &    22.5843 &    23:53:04 &    +14:35:07 &    0.0263 &     107.5 &      Ha5     & 20140825 \\
  332879 &    22.7499 &    23:56:44 &    +15:27:36 &    0.0265 &     108.5 &      Ha5     & 20131010 \\
  332887 &    23.4223 &    23:58:44 &    +16:05:26 &    0.0196 &     83.2  &      Ha4     & 20141018 \\
  332906 &    23.3617 &    23:05:09 &    +25:52:28 &    0.0327 &     135.1 &      Ha5     & 20140101 \\
  333224 &    22.9186 &    23:59:24 &    +26:32:53 &    0.0257 &     105.4 &      Ha4     & 20140101 \\
  333318 &    22.7712 &    23:10:39 &    +24:08:40 &    0.0410  &     170.5 &      Ha6     & 20140827 \\
  333442 &    22.6876 &    23:58:33 &    +31:07:47 &    0.0320  &     132.5 &      Ha5     & 20160108 \\
  748648 &    23.4768 &    21:44:47 &    +15:24:26 &    0.0378 &     157.3 &      Ha6     & 20140827 \\
  748715 &    22.7025 &    22:39:38 &    +13:57:58 &    0.0208 &     89.8  &      Ha4     & 20140825 \\
  748723 &    23.7324 &    22:52:04 &    +15:12:20 &    0.0373 &     154.7 &      Ha6     & 20140827 \\
  748724 &    22.8417 &    22:55:07 &    +14:48:04 &    0.0314 &     129.6 &      Ha5     & 20131228 \\
  748737 &    22.9517 &    23:03:03 &    +14:10:13 &    0.0247 &     100.6 &      Ha4     & 20141018 \\
  748738 &    24.3426 &    23:04:52 &    +14:01:05 &    0.0130  &     56.5  &      Ha3     & 20131231 \\
  748744 &    23.1245 &    23:09:16 &    +14:21:58 &    0.0163 &     70.5  &      Ha3     & 20140819 \\
  748757 &    22.5773 &    23:19:04 &    +16:01:20 &    0.0130  &     56.1  &      Ha3     & 20140819 \\
  748763 &    22.7211 &    23:23:32 &    +13:50:16 &    0.0437 &     182.2 &      Ha7     & 20141017 \\
  748765 &    24.5539 &    23:23:43 &    +14:25:40 &    0.0116 &     50.0  &      Ha3     & 20140826 \\
  748766 &    23.3312 &    23:23:48 &    +14:56:50 &    0.0425 &     177.0 &      Ha7     & 20140102 \\
  748767 &    23.3730  &    23:24:11 &    +15:53:10 &   0.0144 &     61.8  &      Ha3     & 20140826 \\
  748769 &    24.4649 &    23:26:14 &    +15:04:41 &    0.0140  &     60.1  &      Ha3     & 20140826 \\
  748770 &    23.0814 &    23:27:29 &    +14:48:48 &    0.0407 &     169.4 &      Ha6     & 20140827 \\
  748777 &    24.2502 &    00:03:11 &    +15:02:40 &    0.0460  &     192.1 &      Ha7     & 20141017 \\
  748778 &    24.5455 &    00:06:34 &    +15:30:39 &    9.0E-4 &     4.6   &      Ha1     & 20140102 \\
  748786 &    23.7866 &    00:23:06 &    +15:08:21 &    0.0184 &     77.7  &      Ha3     & 20140819 \\
  748788 &    22.8876 &    00:24:10 &    +15:59:38 &    0.0174 &     73.5  &      Ha3     & 20141021 \\
  748790 &    23.1116 &    00:25:07 &    +14:22:06 &    0.0180  &     75.9  &      Ha3     & 20131229 \\
  748794 &    24.1651 &    00:39:28 &    +14:37:07 &    0.0177 &     74.3  &      Ha3     & 20131228 \\
  748795 &    22.5902 &    00:40:56 &    +14:14:08 &    0.0387 &     161.1 &      Ha6     & 20160209 \\
  748798 &    24.6805 &    00:49:01 &    +14:03:05 &    0.0386 &     160.6 &      Ha6     & 20160209 \\
  748805 &    22.8382 &    01:04:36 &    +15:16:21 &    0.0144 &     59.6  &      Ha3     & 20141021 \\
  748815 &    22.6130  &    01:27:03 &    +14:39:38 &   0.0216 &     88.0  &      Ha4     & 20140825 \\
  748817 &    22.5455 &    01:28:33 &    +15:14:54 &    0.0211 &     86.2  &      Ha4     & 20141018 \\
  748819 &    24.6432 &    01:37:25 &    +14:39:37 &    0.0086 &     35.0  &      Ha2     & 20160207 \\
\enddata
\end{deluxetable*}

\startlongtable
\begin{deluxetable*}{rrcccccrc}
\tabletypesize{\scriptsize}
\tablecolumns{9}
\tablewidth{-20pt}
\tablecaption{The Star Formation Properties of LSBGs \label{table3}}
\tabletypesize{\scriptsize}
\tablehead { 
\colhead{AGC}&  \colhead{$r_{25}$} & \colhead{Ellipse} & \colhead{$\rm log F(H\alpha)$} & \colhead{$\rm log \sigma( F(H\alpha))$}&\colhead{log$\rm (SFR)$}&\colhead{$\rm log \Sigma_{sfr}$}&\colhead{$\rm log M_{HI}$}&\colhead{$\rm log \Sigma_{HI}$} \\
\colhead{}&  \colhead{Kpc}&\colhead{}& \colhead{$erg cm^{-2} s^{-1}$} &  \colhead{$erg cm^{-2} s^{-1}$} & \colhead{$M_{\odot}yr^{-1}$} & \colhead{$M_{\odot}yr^{-1}Kpc^{-2}$}& \colhead{$M_{\odot}$}&  \colhead{$M_{\odot}pc^{-2}$}\\
\colhead{(1)} &\colhead{(2)} &\colhead{(3)} &\colhead{(4)} &\colhead{(5)} &\colhead{(6)} &\colhead{(7)} &\colhead{(8)}&\colhead{(9)} }
\startdata
             17    &      2.63    &      0.24    &  \nodata    &  \nodata    &  \nodata    &  \nodata    &      8.41    &      0.73   \\
            273    &      9.40    &      0.32    &    -14.64    &      0.08    &     -1.87    &     -4.15    &      9.59    &      0.85   \\
            337    &      8.47    &      0.26    &    -12.93    &      0.01   &     -0.20    &     -2.43    &      9.88    &      1.20   \\
           1084    &      5.67    &      0.47    &    -13.78    &      0.02    &     -1.47    &     -3.19    &      9.50    &      1.31   \\
           1211    &      6.89    &      0.33    &    -13.30    &      0.01    &     -1.30    &     -3.30    &      8.98    &      0.52   \\
           1362    &     12.44    &      0.23    &    -13.73    &      0.01    &     -0.68    &     -3.25    &      9.36    &      0.33   \\
           1693    &      9.27    &      0.07    &    -13.22    &      0.01    &     -0.81    &     -3.21    &      9.49    &      0.63   \\
           2144    &      9.66    &      0.20    &    -12.96    &      0.01    &     -0.35    &     -2.72    &      9.18    &      0.35   \\
          12289    &     24.96    &      0.11    &    -13.17    &      0.01    &      0.10    &     -3.14    &     10.30    &      0.60   \\
          12845    &     25.03    &      0.20    &    -11.67    &      0.01    &      0.98    &     -2.22    &     10.18    &      0.52   \\
         100037    &      6.07    &      0.20    &    -13.35    &      0.01    &     -1.07    &     -3.04    &      8.76    &      0.33   \\
         100350    &      5.27    &      0.20    &    -14.23    &      0.04    &     -1.63    &     -3.47    &      8.92    &      0.62   \\
         101191    &      5.83    &      0.33    &    -13.44    &      0.01    &     -0.72    &     -2.57    &      8.95    &      0.63   \\
         101812    &      1.89    &      0.20    &    -13.94    &      0.02    &     -2.11    &     -3.06    &      8.73    &      1.31   \\
         101877    &      7.87    &      0.50    &    -13.97    &      0.01    &     -1.27    &     -3.26    &      9.57    &      1.12   \\
         101942    &      5.68    &      0.45    &    -14.96    &      0.10    &     -2.18    &     -3.93    &      9.14    &      0.93   \\
         101986    &      8.54    &      0.20    &    -13.70    &      0.01    &     -0.69    &     -2.95    &      9.33    &      0.61   \\
         102098    &      9.30    &      0.34    &  \nodata    &  \nodata    &  \nodata    &  \nodata    &      9.68    &      0.97   \\
         102101    &      8.39    &      0.35    &    -13.78    &      0.02    &     -1.04    &     -3.20    &      9.21    &      0.59   \\
         102229    &      9.68    &      0.20    &  \nodata    &  \nodata    &  \nodata    &  \nodata    &      8.94    &      0.11   \\
         102243    &      8.99    &      0.24    &    -13.52    &      0.01    &     -0.65    &     -2.93    &      9.78    &      1.03   \\
         102302    &      2.04    &      0.20    &    -15.01    &      0.18    &     -3.22    &     -4.24    &      8.79    &      1.31   \\
         102558    &      8.66    &      0.03    &  \nodata    &  \nodata    &  \nodata    &  \nodata    &      8.27    &     -0.55   \\
         102630    &      5.89    &      0.20    &    -14.11    &      0.05    &     -1.24    &     -3.18    &      9.17    &      0.77   \\
         102635    &     12.82    &      0.14    &    -13.56    &      0.01    &     -0.35    &     -3.00    &      9.65    &      0.54   \\
         102672    &      4.19    &      0.29    &    -13.16    &      0.01    &     -0.44    &     -2.04    &      9.20    &      1.15   \\
         102674    &     13.94    &      0.12    &    -13.97    &      0.01    &     -0.42    &     -3.15    &     10.02    &      0.83   \\
         102684    &      8.57    &      0.43    &    -13.80    &      0.04    &     -0.81    &     -2.93    &      9.27    &      0.69   \\
         102728    &      0.23    &      0.20    &    -14.47    &      0.06    &     -3.58    &     -2.70    &      6.78    &      1.20   \\
         102729    &      4.81    &      0.34    &    -13.59    &      0.01    &     -0.99    &     -2.67    &      8.85    &      0.71   \\
         102730    &     10.12    &      0.22    &    -14.00    &      0.01    &     -0.54    &     -2.94    &      9.68    &      0.82   \\
         102900    &     16.24    &      0.20    &    -14.02    &      0.03    &     -0.59    &     -3.41    &      9.81    &      0.53   \\
         102981    &      7.72    &      0.20    &  \nodata    &  \nodata    &  \nodata    &  \nodata    &      8.72    &      0.08   \\
         110150    &      6.78    &      0.05    &    -13.33    &      0.01    &     -0.97    &     -3.10    &      9.49    &      0.89   \\
         110319    &      5.55    &      0.20    &    -14.13    &      0.02    &     -1.46    &     -3.35    &      9.22    &      0.87   \\
         110379    &      2.72    &      0.16    &    -13.88    &      0.01    &     -1.86    &     -3.15    &      9.20    &      1.45   \\
         110398    &     12.53    &      0.42    &    -13.40    &      0.01    &     -0.50    &     -2.95    &      9.63    &      0.71   \\
         112503    &      1.22    &      0.55    &    -13.42    &      0.01    &     -2.43    &     -2.75    &      7.14    &      0.36   \\
         112892    &     10.28    &      0.20    &    -13.79    &      0.05    &     -0.44    &     -2.86    &      9.54    &      0.66   \\
         113200    &      5.18    &      0.20    &    -13.91    &      0.06    &     -0.92    &     -2.75    &      9.29    &      1.00   \\
         113752    &     17.34    &      0.20    &  \nodata    &  \nodata    &  \nodata    &  \nodata    &      9.72    &      0.38   \\
         113790    &      5.30    &      0.20    &  \nodata    &  \nodata    &  \nodata    &  \nodata    &      8.57    &      0.26   \\
         113825    &      2.13    &      0.20    &    -14.66    &      0.05    &     -2.24    &     -3.30    &      8.98    &      1.46   \\
         113845    &      4.70    &      0.28    &    -14.43    &      0.02    &     -1.35    &     -3.05    &      9.25    &      1.09   \\
         113907    &      8.61    &      0.10    &    -13.96    &      0.02    &     -0.67    &     -3.00    &      9.36    &      0.58   \\
         113918    &      9.87    &      0.11    &  \nodata    &  \nodata    &  \nodata    &  \nodata    &      9.47    &      0.57   \\
         113923    &      3.34    &      0.20    &    -14.12    &      0.01    &     -1.63    &     -3.08    &      9.05    &      1.14   \\
         114040    &      8.07    &      0.42    &    -14.83    &      0.34    &     -1.79    &     -3.86    &      9.40    &      0.86   \\
         121174    &      0.88    &      0.20    &  \nodata    &  \nodata    &  \nodata    &  \nodata    &      8.18    &      1.43   \\
         122138    &      1.25    &      0.20    &    -14.25    &      0.04    &     -3.00    &     -3.59    &      8.08    &      1.02   \\
         122210    &      6.27    &      0.23    &    -13.46    &      0.01    &     -0.90    &     -2.88    &      9.29    &      0.85   \\
         122211    &      3.97    &      0.32    &    -13.96    &      0.02    &     -1.59    &     -3.11    &      8.66    &      0.67   \\
         122341    &     14.42    &      0.45    &    -13.95    &      0.03    &     -0.59    &     -3.14    &      9.90    &      0.88   \\
         122874    &      5.05    &      0.37    &    -14.29    &      0.02    &     -1.43    &     -3.13    &      9.12    &      0.96   \\
         122877    &      8.55    &      0.80    &  \nodata    &  \nodata    &  \nodata    &  \nodata    &      9.18    &      1.06   \\
         122884    &      2.93    &      0.33    &    -14.03    &      0.02    &     -2.03    &     -3.29    &      9.17    &      1.45   \\
         122924    &      4.05    &      0.19    &  \nodata    &  \nodata    &  \nodata    &  \nodata    &      9.50    &      1.42   \\
         123046    &      8.51    &      0.07    &    -13.30    &      0.03    &     -0.69    &     -3.01    &      8.89    &      0.10   \\
         123047    &      6.42    &      0.45    &    -13.98    &      0.02    &     -0.69    &     -2.55    &      9.47    &      1.16   \\
         123170    &      1.29    &      0.20    &  \nodata    &  \nodata    &  \nodata    &  \nodata    &      7.68    &      0.60   \\
         123172    &      5.77    &      0.30    &    -13.96    &      0.02    &     -1.24    &     -3.10    &      9.25    &      0.92   \\
         320466    &      5.71    &      0.35    &  \nodata    &  \nodata    &  \nodata    &  \nodata    &      9.13    &      0.85   \\
         321166    &      6.48    &      0.20    &  \nodata    &  \nodata    &  \nodata    &  \nodata    &      8.62    &      0.14   \\
         321341    &     11.90    &      0.26    &    -14.13    &      0.03    &     -0.70    &     -3.22    &      9.66    &      0.68   \\
         321348    &      9.64    &      0.31    &    -13.73    &      0.01    &     -0.53    &     -2.84    &      9.52    &      0.75   \\
         321385    &      3.95    &      0.35    &    -14.19    &      0.02    &     -1.23    &     -2.73    &      9.32    &      1.36   \\
         321429    &      1.74    &      0.56    &    -15.53    &      0.20    &     -3.06    &     -3.69    &      8.55    &      1.47   \\
         321435    &      4.74    &      0.61    &    -13.29    &      0.01    &     -0.81    &     -2.25    &      8.88    &      0.98   \\
         321438    &      4.32    &      0.39    &    -15.05    &      0.12    &     -2.01    &     -3.56    &      9.12    &      1.11   \\
         321451    &      6.90    &      0.20    &    -13.44    &      0.01    &     -0.50    &     -2.57    &      9.31    &      0.77   \\
         321490    &      6.79    &      0.45    &    -14.02    &      0.01    &     -1.09    &     -2.99    &      9.37    &      1.01   \\
         321492    &      5.00    &      0.19    &  \nodata    &  \nodata    &  \nodata    &  \nodata    &      7.73    &     -0.53   \\
         331052    &      4.74    &      0.09    &    -13.65    &      0.01    &     -1.03    &     -2.83    &      8.78    &      0.51   \\
         332431    &      9.61    &      0.29    &    -13.50    &      0.01    &     -0.53    &     -2.84    &      9.64    &      0.86   \\
         332640    &      9.60    &      0.29    &    -14.35    &      0.03    &     -1.30    &     -3.61    &      9.62    &      0.85   \\
         332761    &      3.88    &      0.03    &    -14.41    &      0.02    &     -1.60    &     -3.27    &      8.57    &      0.45   \\
         332786    &     12.10    &      0.51    &  \nodata    &  \nodata    &  \nodata    &  \nodata    &      8.57    &     -0.24   \\
         332844    &     10.54    &      0.26    &    -13.98    &      0.02    &     -0.57    &     -2.99    &      9.64    &      0.77   \\
         332861    &     10.48    &      0.54    &    -14.64    &      0.04    &     -1.60    &     -3.80    &      9.40    &      0.74   \\
         332879    &      6.73    &      0.20    &    -14.36    &      0.05    &     -1.32    &     -3.37    &      9.31    &      0.79   \\
         332887    &      4.65    &      0.17    &    -14.21    &      0.01    &     -1.40    &     -3.15    &      9.19    &      0.98   \\
         332906    &      7.65    &      0.08    &    -14.20    &      0.04    &     -0.96    &     -3.19    &      9.58    &      0.89   \\
         333224    &      9.57    &      0.58    &    -14.39    &      0.06    &     -1.37    &     -3.46    &      9.50    &      0.96   \\
         333318    &     17.89    &      0.18    &    -14.02    &      0.06    &     -0.58    &     -3.49    &      9.95    &      0.57   \\
         333442    &     13.26    &      0.32    &    -14.04    &      0.02    &     -0.82    &     -3.39    &      9.65    &      0.61   \\
         748648    &      5.42    &      0.35    &    -14.54    &      0.03    &     -1.17    &     -2.95    &      9.69    &      1.45   \\
         748715    &      3.03    &      0.38    &    -15.13    &      0.07    &     -2.25    &     -3.50    &      9.16    &      1.45   \\
         748723    &     17.70    &      0.70    &    -13.84    &      0.03    &     -0.48    &     -2.95    &      9.61    &      0.68   \\
         748724    &      9.63    &      0.20    &    -14.10    &      0.03    &     -0.90    &     -3.27    &      9.60    &      0.77   \\
         748737    &      5.83    &      0.20    &    -14.85    &      0.11    &     -1.87    &     -3.80    &      9.54    &      1.15   \\
         748738    &      2.27    &      0.26    &    -14.76    &      0.04    &     -2.28    &     -3.36    &      8.61    &      1.07   \\
         748744    &      3.44    &      0.46    &    -14.52    &      0.02    &     -1.85    &     -3.15    &      9.02    &      1.26   \\
         748757    &      5.14    &      0.58    &    -13.80    &      0.01    &     -1.32    &     -2.86    &      9.32    &      1.32   \\
         748763    &      5.44    &      0.16    &    -14.73    &      0.03    &     -1.24    &     -3.13    &      9.90    &      1.55   \\
         748765    &      2.24    &      0.20    &    -12.72    &      0.01    &     -0.35    &     -1.45    &      8.63    &      1.07   \\
         748766    &     10.21    &      0.33    &    -14.40    &      0.08    &     -0.93    &     -3.27    &      9.89    &      1.09   \\
         748767    &      3.56    &      0.08    &    -15.30    &      0.14    &     -2.74    &     -4.31    &      8.84    &      0.81   \\
         748769    &      3.47    &      0.20    &    -14.51    &      0.04    &     -1.97    &     -3.45    &      8.76    &      0.82   \\
         748770    &     11.99    &      0.35    &    -14.31    &      0.07    &     -0.88    &     -3.34    &      9.71    &      0.78   \\
         748777    &     13.85    &      0.41    &    -14.12    &      0.02    &     -0.58    &     -3.13    &      9.88    &      0.87   \\
         748778    &      0.15    &      0.20    &    -13.99    &      0.04    &     -3.69    &     -2.46    &      6.36    &      1.13   \\
         748786    &      7.91    &      0.78    &    -13.87    &      0.01    &     -1.11    &     -2.75    &      9.15    &      1.05   \\
         748788    &      2.87    &      0.05    &    -13.66    &      0.01    &     -0.95    &     -2.34    &      8.85    &      1.00   \\
         748790    &      4.78    &      0.70    &  \nodata    &  \nodata    &  \nodata    &  \nodata    &      8.82    &      1.03   \\
         748794    &      4.70    &      0.20    &  \nodata    &  \nodata    &  \nodata    &  \nodata    &      9.13    &      0.93   \\
         748795    &     10.04    &      0.48    &    -14.18    &      0.04    &     -0.79    &     -3.01    &      9.64    &      0.96   \\
         748798    &      6.55    &      0.20    &    -15.40    &      0.34    &     -2.01    &     -4.05    &      9.55    &      1.06   \\
         748805    &      8.71    &      0.17    &    -14.77    &      0.08    &     -2.25    &     -4.54    &      8.74    &     -0.02   \\
         748815    &      4.13    &      0.39    &    -14.03    &      0.01    &     -1.17    &     -2.68    &      9.23    &      1.25   \\
         748817    &      4.82    &      0.29    &    -14.18    &      0.02    &     -1.33    &     -3.04    &      9.17    &      0.99   \\
         748819    &      6.39    &      0.20    &  \nodata    &  \nodata    &  \nodata    &  \nodata    &      8.78    &      0.31   \\ 
\enddata
\end{deluxetable*}
\bibliographystyle{aasjournal}
\bibliography{sample}

\end{document}